\documentclass[nhess, manuscript]{copernicus}

\begin{document}

\title{Scaling-law-informed neural point processes for earthquake sequence forecasting}

\Author[1,2]{Tianlu}{Xiong}
\Author[1,2]{Zaibo}{Zhao}
\Author[1,2]{Yunrui}{Li}
\Author[1,2]{Wenqi}{Liu}
\Author[3]{Yosef}{Ashkenazy}
\Author[1,2][zhangyongwen77@gmail.com]{Yongwen}{Zhang}

\affil[1]{Yunnan Key Laboratory of Complex Systems and Brain-Inspired Intelligence,
Kunming University of Science and Technology, Kunming, Yunnan, China}

\affil[2]{Faculty of Science, Kunming University of Science and Technology,
Kunming, Yunnan, China}

\affil[3]{Environmental Physics, Ben-Gurion University of the Negev,
Midreshet Ben-Gurion, Israel}

\runningtitle{Scaling-law-informed neural earthquake forecasting}

\runningauthor{T. Xiong et al.}

\received{}
\pubdiscuss{} 
\revised{}
\accepted{}
\published{}

\firstpage{1}

\maketitle

\nolinenumbers
\renewcommand{\baselinestretch}{1}\normalsize

\begin{abstract}
Earthquake sequence forecasting requires models that can learn nonlinear history dependence while retaining robust statistical structure. We develop a scaling-law-informed neural marked point process, termed Fusion, that combines neural representations of catalog history with temporal features derived from the Epidemic-Type Aftershock Sequence (ETAS) model and magnitude information derived from the Gutenberg--Richter law. The model separates the magnitude cutoff applied to the input catalog from the fixed target-event threshold, allowing lower-magnitude earthquakes to inform forecasts without changing the target-event set. For the 2016--2017 Amatrice--Visso--Norcia sequence, Fusion achieves the highest target-event temporal likelihood when lower-magnitude events are retained, outperforming both the ETAS model and a purely neural point-process baseline. Event-wise and cumulative analyses show sustained timing gains through substantial portions of the Visso and Norcia sequences. Across five benchmark catalogs, catalog-specific neural training with a fixed ETAS prior yields the highest temporal likelihood at the minimum evaluated magnitude cutoff. Magnitude likelihood shows no consistent predictive gain beyond the Gutenberg--Richter-based ETAS reference, indicating that the additional information captured by Fusion is primarily temporal. These results show that lower-magnitude catalog histories and empirical scaling-law information complement neural sequence learning for target-event timing.
\end{abstract}


\introduction  
Earthquake sequence forecasting is inherently history dependent: each event can alter the short-term probability of subsequent events, and aftershock triggering can dominate seismicity rates following large earthquakes~\citep{ref19}. Catalogs therefore contain clustered and evolving sequences rather than independent stationary samples, as illustrated by the well-documented 2016--2017 Amatrice--Norcia sequence~\citep{ref31}. Point-process models provide a natural probabilistic framework for this setting~\citep{ref36,ref37}. Common likelihood-based evaluation protocols enable comparisons across models and sequences in operational forecasting~\citep{ref27,ref32}. The central modeling challenge is thus to combine adaptive sequence learning with seismological interpretability and reliable out-of-sample performance~\citep{ref35}.

The Epidemic-Type Aftershock Sequence (ETAS) model is the principal statistical reference for this problem because it represents seismicity as a self-exciting point process with a background rate and earthquake-triggered aftershock contributions~\citep{ref36,ref37,ref38}. It embeds the Gutenberg--Richter magnitude-frequency law~\citep{ref17} and the Omori--Utsu law for temporal aftershock decay~\citep{ref47,ref48}, providing a clear seismological interpretation and a strong reference for short-term forecasts~\citep{ref19}. Forecast performance nevertheless depends on catalog truncation, calibration assumptions, triggering cascades and parameter uncertainty~\citep{ref42,ref65,ref70,ref71}. Fixed parametric assumptions can also become restrictive when catalogs are incomplete or their completeness changes~\citep{ref33,ref39,ref50}, magnitude-frequency behavior evolves~\citep{ref16,ref26}, or temporal dependencies extend beyond simple kernels~\citep{ref58,ref59,ref60}.

These catalog effects make the input magnitude cutoff more than a preprocessing choice. Small and moderate earthquakes may fall below the forecasting target but can still provide a denser record of evolving aftershock activity and useful historical information for forecasting larger target events~\citep{ref19,ref44}. Removing them can discard useful historical context, whereas retaining them increases sensitivity to completeness limits and catalog truncation~\citep{ref33,ref42,ref50}. A forecasting experiment must therefore distinguish the events available as model history from those scored as targets. Following the neural point-process protocol developed for the 2016--2017 Central Apennines sequence~\citep{ref44}, we separate the magnitude threshold defining the retained input history from the fixed threshold defining the target-event set.

Neural point processes offer a complementary way to model earthquake sequences. By learning conditional intensities and marked-event distributions from event histories, they can represent nonlinear dependencies that are difficult to specify with closed-form triggering functions. They have been applied to earthquake sequence and rate forecasting~\citep{ref10,ref44,ref57}, while related deep-learning and multimodal approaches show broader potential across regional and induced-seismicity settings~\citep{ref22,ref41,ref56}. Cross-catalog benchmarks now enable direct comparisons between neural point processes and ETAS across magnitude thresholds and show that purely neural models do not consistently outperform ETAS~\citep{ref69}. Existing neural point-process models generally do not explicitly encode ETAS triggering or Gutenberg--Richter scaling within their conditional likelihoods~\citep{ref44,ref69}.

This tension motivates embedding seismological knowledge within neural forecasting architectures. Scientific machine learning shows how domain structure can guide data-driven models~\citep{ref12,ref28,ref49}, while Earth-system artificial intelligence (AI) research emphasizes hybrid modeling and explainability for trustworthy geoscience applications~\citep{ref08,ref09,ref11,ref25}. In earthquake forecasting, the most direct prior knowledge often comes from empirical seismicity laws and ETAS-type triggering rather than governing dynamic equations. ETAS-inspired neural and hybrid models have begun to combine data-driven learning with aftershock-triggering structure~\citep{ref53,ref54}, and recent reviews and seismically informed AI studies further highlight the value of incorporating seismological priors~\citep{ref55,ref63}. These studies establish the promise of hybrid forecasting. Within this emerging literature, however, limited attention has been given to likelihood-based marked point processes that jointly use ETAS-derived temporal quantities and Gutenberg--Richter magnitude information as internal features.

We therefore develop a scaling-law-informed neural marked point process for earthquake sequence forecasting. The framework uses ETAS-derived triggering and Gutenberg--Richter magnitude quantities as structured prior inputs rather than strict constraints, while retaining the flexibility of neural sequence learning. It explicitly separates the retained catalog history from the evaluated target-event set. On the Amatrice--Visso--Norcia sequence, we assess how lower-magnitude historical events affect the timing forecasts of a fixed target set; the magnitude likelihood provides a corresponding test of whether neural conditioning adds predictive information beyond the Gutenberg--Richter distribution. We then evaluate performance consistency across five EarthquakeNPP benchmark catalogs using catalog-specific neural training and a fixed ETAS feature source.

\section{Methods}\label{sec:method}
\subsection{Catalog datasets and preprocessing}\label{subsec:preprocess}

We use two groups of earthquake catalogs. The primary catalog is the 2016--2017 Amatrice--Visso--Norcia (AVN) earthquake sequence in the central Apennines, Italy~\citep{ref44}, covering the spatial range \(42.1985^\circ\)--\(43.9180^\circ\)~N and \(11.6882^\circ\)--\(13.8239^\circ\)~E. The AVN catalog was originally developed by ~\citet{ref72} and was obtained in the processed form provided by ~\citet{ref44}. This dataset is used in Figs.~\ref{fig1:framework}a, b and ~\ref{fig2:LL}--\ref{fig4:tig_PDF}, while Fig.~\ref{fig1:framework}c presents only the schematic architecture of Fusion. We use the processed time--magnitude event sequence adopted in the neural point-process benchmark of ~\citet{ref44}, so that ETAS, the Neural Point-Process (NPP) baseline and Fusion are compared under the same catalog construction. The AVN sequence is evaluated using three training--testing splits associated with the Visso, Norcia and Campotosto forecasting windows. The corresponding split points are defined by event indices 1200, 1800 and 3600 in the raw catalog, respectively. Each index determines a fixed chronological split time that is retained for every \(M_{\mathrm{cut}}\): events before that time are used for training and events after it are used for testing. For neural models, the last 20 retained training events are appended to the beginning of the corresponding test sequence as burn-in context before the first scored test forecast.

To assess performance across catalogs, we also use the five EarthquakeNPP benchmark catalogs~\citep{ref69}: Comprehensive Earthquake Catalog (ComCat), QTM-SaltonSea (QTM-SS), QTM-SanJac (QTM-SJ), Southern California Earthquake Data Center (SCEDC) and WHITE. The raw catalogs underlying these benchmarks are Advanced National Seismic System (ANSS) ComCat maintained by the U.S. Geological Survey (USGS)~\citep{ref73}, SCEDC provided by SCEDC/	Southern California Seismic Network (SCSN)~\citep{ref74, ref75}, the Quake Template Matching (QTM) catalog developed by ~\citet{ref39}, and the San Jacinto fault-zone catalog developed by ~\citet{ref77}. These benchmark catalogs mainly cover California and Southern California seismic regions. The five benchmark datasets used in Figs.~\ref{fig5:benchmark_box}a--e were obtained from the EarthquakeNPP project~\citep{ref69}, whereas Fig.~\ref{fig5:benchmark_box}f uses the Campotosto split of the AVN dataset. The five EarthquakeNPP benchmark catalogs also differ substantially in spatial coverage. ComCat spans the broadest regional domain, whereas QTM-SS, QTM-SJ and WHITE are more localized southern California catalogs; the detailed longitude--latitude ranges of the benchmark catalogs are provided in Appendix~\ref{app:catalog-domains}.
We follow the official benchmark partition: the Auxiliary and Training periods are used as training history, while the validation and testing periods follow the official setting. The official minimum input cutoff thresholds are $M_{\mathrm{cut}}=2.5$ for ComCat, $1.0$ for QTM-SaltonSea, $1.0$ for QTM-SanJac, $2.0$ for SCEDC and $0.6$ for WHITE. Each benchmark catalog is also evaluated at the common cutoff $M_{\mathrm{cut}}=3.0$.

For all catalogs, each earthquake event is represented only by its occurrence time and magnitude, denoted as \((t_i,m_i)\); event locations, including latitude and longitude, are used only to document the catalog spatial extent and are not included as model inputs. AVN magnitudes are reported as moment magnitude $M_w$, whereas $m_i$ for the EarthquakeNPP datasets denotes the catalog-reported magnitude supplied by the benchmark rather than a uniformly converted magnitude type. For a given input magnitude threshold $M_{\mathrm{cut}}$, events satisfying $m_i \geq M_{\mathrm{cut}}$ are retained to construct the input catalog,
\begin{equation}
\mathcal{S}
=
\{(t_i,m_i)\}_{i=1}^{N},
\qquad
0<t_1<\cdots<t_N .
\label{eq:catalog-sequence}
\end{equation}

Throughout the formulation, $i$ indexes the current retained event or retained-event interval, $j$ indexes previous retained events inside histories or triggering sums, and $n$ indexes the target-event subsequence.
The retained history before a candidate time $t$ is

\[
H_t
=
\{(t_j,m_j):t_j<t,\;m_j\geq M_{\mathrm{cut}}\}.
\]

Thus, when the candidate time is the retained event time $t_i$, we write the available history as $H_i=H_{t_i}=\{(t_j,m_j):j<i\}$. The observed elapsed time between two consecutive retained events is

\begin{equation}
\Delta_i=t_i-t_{i-1},
\label{eq:retained-interevent-time}
\end{equation}

with $t_0$ denoting the start of the evaluated sequence. For a candidate time $t$ after $t_{i-1}$, we use the candidate elapsed time $\delta=t-t_{i-1}>0$; setting $t=t_i$ gives $\delta=\Delta_i$. This retained sequence, rather than the target-event subsequence alone, is used as the conditioning history for ETAS, NPP and Fusion.

The forecasting target threshold is fixed at $M_d=3.0$ and is introduced in the likelihood and evaluation protocol below. All cutoff settings considered in this study satisfy $M_{\mathrm{cut}}\leq M_d$, ensuring that every target event remains in the retained catalog. When $M_{\mathrm{cut}} < M_d$, lower-magnitude retained events can still enter the model history and temporal exposure, but they are not treated as event-occurrence targets in the likelihood.

\subsection{ETAS baseline}\label{subsec:etas}

The ETAS model is used as the statistical seismicity baseline. 
ETAS represents earthquake occurrence as a self-exciting point process, in which the conditional occurrence rate is decomposed into a background seismicity component and a triggering component induced by previous events. 
For a given history $H_t$, the fitted ETAS temporal intensity is written as
\begin{equation}
\lambda_0^{\mathrm{ETAS}}(t\mid H_t)
=
\mu
+
\sum_{t_j<t}
k(m_j)\,g(t-t_j),
\label{eq:etas-intensity}
\end{equation}

where $\mu$ is the background rate, $k(m_j)$ is the productivity function of the previous event with magnitude $m_j$, and $g(\cdot)$ is the temporal triggering kernel.

Here $k(m_j)=k_0\exp[a(m_j-M_{\mathrm{cut}})]$ is the productivity term for an ETAS fit estimated at input cutoff $M_{\mathrm{cut}}$, and $g(\cdot)$ follows the Omori--Utsu aftershock-decay kernel. 
The standard ETAS formulation is used as the statistical reference because it provides an interpretable and reproducible baseline based on established empirical seismicity laws, without introducing additional assumptions from ETAS variants.
The magnitude component follows the Gutenberg--Richter law with a finite upper bound $M_{\max}=8.0$, which is used for analytical normalization in ETAS and as the effective evaluation range for the neural models. The closed-form cumulative temporal intensity, the truncated Gutenberg--Richter density and the target-conditioned ETAS quantities are listed in Appendix~\ref{app:target-etas}.

The ETAS parameters are estimated by maximum likelihood on the retained catalog specified by the corresponding experimental setting. 
Because the fitted ETAS process describes the retained catalog above the input cutoff \(M_{\mathrm{cut}}\), whereas the likelihood evaluation is performed only on target events with \(m\geq M_d=3.0\), we convert the ETAS temporal intensity and magnitude distribution to their corresponding target-threshold forms before likelihood evaluation. This conditioning is part of the common likelihood protocol described below, and its derivation is given in Appendix~\ref{app:target-etas}.

\subsection{Neural point-process baseline}\label{subsec:neural-models}\label{subsec:npp}

The Neural Point Process (NPP) baseline is the temporal--magnitude neural point-process model introduced by ~\citet{ref44}, rather than a generic NPP implementation. Unlike ETAS, NPP does not impose an Omori--Utsu decay kernel or Gutenberg--Richter magnitude law; it learns cumulative temporal and magnitude functions directly from earthquake sequences. NPP and Fusion use the same likelihood protocol introduced below, but NPP constructs its representation only from the Long Short-Term Memory (LSTM)-based history representation.

For each retained event, we construct a normalized input vector from the retained-event inter-event time and magnitude. The normalized input vector is defined as

\begin{equation}
\boldsymbol{x}_i
=
\begin{bmatrix}
(\log \Delta_i-\mu_t)/\sigma_t \\
(\log m_i-\mu_m)/\sigma_m
\end{bmatrix},
\end{equation}

where $\varepsilon_t=10^{-10}$ is added for numerical stability, and $(\mu_t,\sigma_t)$ and $(\mu_m,\sigma_m)$ are the mean and standard deviation of $\log(\Delta+\varepsilon_t)$ and $\log m$, respectively, computed from the training data. The candidate elapsed time supplied to the temporal and magnitude networks of both NPP and Fusion is transformed in the same way, as $[\log(\delta+\varepsilon_t)-\mu_t]/\sigma_t$; $\phi_{\Delta}(\delta)$ denotes its embedding in the Fusion equations. No additional clipping or special treatment of candidate elapsed times is applied.
The logarithmic transformation is used as a neural-network input preprocessing step rather than as a physical redefinition of magnitude. For the inter-event time \(\Delta_i\), this transformation is important because earthquake waiting times are highly uneven and can span several orders of magnitude: events are densely clustered after major earthquakes, whereas inter-event times can become much longer during quieter periods. Directly using raw \(\Delta_i\) would make large waiting times dominate the input scale and could weaken the LSTM's ability to learn short-time triggering structure. Using \(\log(\Delta_i+\varepsilon_t)\) compresses this dynamic range and allows the recurrent encoder to learn relative temporal variations more stably. Although earthquake magnitude is itself a logarithmic measure of seismic moment, the use of \(\log m_i\) here serves a different purpose. The magnitude input to the LSTM is a numerical covariate whose empirical distribution is strongly imbalanced, with many more small-to-moderate events than large events. Applying \(\log m_i\), followed by standardization, slightly compresses the magnitude range and reduces scale imbalance between the time and magnitude inputs. This preprocessing improves the numerical stability of recurrent-history learning. In sensitivity experiments, replacing \(\log m_i\) with the raw magnitude \(m_i\) led to worse predictive performance, so we retain the logarithmic magnitude input for the NPP and Fusion history encoders.

To avoid instability caused by extremely long recurrent histories, we use a fixed history window of length $L=20$. The history representation before the current event is obtained by an LSTM:

\begin{equation}
\boldsymbol{h}_i
=
\mathrm{LSTM}\!\left(\boldsymbol{x}_{i-L},\ldots,\boldsymbol{x}_{i-1}\right),
\end{equation}

where $\boldsymbol{h}_i$ encodes the recent time--magnitude history before the current event.
The fixed event-count window provides a stable recurrent input length. It is applied in a sliding manner along the retained sequence, so the model is trained and evaluated over all retained events, while the recurrent representation for each forecast conditions only on the most recent \(L\) retained events; events preceding this window do not enter \(\boldsymbol{h}_i\) directly. We use an event-based window instead of a fixed-duration time window because seismicity rates vary strongly across the sequence: a fixed time window could contain very many events immediately after a major earthquake but few or no events during quieter periods, leading to highly variable input sizes. In contrast, an event-count window keeps the neural input dimension fixed, while the inter-event times \(\Delta_i\) still encode the actual temporal spacing between events. The choice \(L=20\) follows the window size used in previous neural point-process earthquake forecasting studies and reflects a practical balance: shorter windows may miss dependencies among recent events, whereas substantially longer windows increase recurrent-history instability and computational cost without clear performance gains in our experiments.

Given the neural history representation $\boldsymbol{h}_i$, NPP parameterizes two cumulative functions over candidate arguments: the temporal cumulative hazard function $\Lambda_i^{\mathrm{NPP}}(\delta\mid \boldsymbol{h}_i)$ for candidate elapsed time $\delta$, and the cumulative magnitude distribution $F_i^{\mathrm{NPP}}(m\mid \delta,\boldsymbol{h}_i)$ for a candidate magnitude $m$. Following ~\citet{ref44}, trainable weights along the cumulative hazard function network (CHFN) paths that depend on $\delta$ and the cumulative magnitude function network (CMFN) paths that depend on $m$ are constrained to be non-negative, and these paths use non-decreasing activations. The CHFN output uses a softplus activation, whereas the CMFN output uses a sigmoid activation. Together these choices make the cumulative outputs non-decreasing in the corresponding candidate variable and yield non-negative derivatives. The LSTM history encoder is built from the retained catalog; the target-event mask is applied later in the likelihood rather than during construction of the recurrent input sequence. The observed values $\Delta_i$ and $m_i$ are substituted only when evaluating the likelihood.

\subsection{Scaling-law-informed Fusion model}\label{subsec:fusion}

The proposed Fusion model (Fig.~\ref{fig1:framework}c) embeds ETAS-derived scaling-law prior information into the neural marked point-process architecture. Unlike NPP, which uses the LSTM history embedding in both components, Fusion uses a branch-specific design: the temporal branch incorporates the LSTM history representation, whereas the magnitude branch does not directly use the LSTM hidden state.

For each retained interval $i$, the temporal branch uses the LSTM history representation $\boldsymbol{h}_i$ and the ETAS-derived temporal feature $I_i(\delta)$. Here ETAS is used not as a final predictor but as a source of scaling-law-informed prior features. The scalar $I_i(\delta)$ is the standardized ETAS temporal integral from $t_{i-1}$ to $t_{i-1}+\delta$, computed from the ETAS feature history specified by the Fusion variant. Fusion VP denotes the variant with varying ETAS parameters, which are refitted at each input magnitude cutoff. The magnitude scalar $G_i(m)$ is the standardized ETAS/Gutenberg--Richter cumulative magnitude quantity at candidate magnitude $m$: Fusion VP (varying ETAS parameters) uses the current-cutoff Gutenberg--Richter distribution estimated at that $M_{\mathrm{cut}}$, whereas Fusion fixed uses the current-cutoff distribution computed with fixed $\beta_{3.0}$. The standardized scalars are mapped by one-layer embeddings $\phi_t(I_i(\delta))$ and $\phi_m(G_i(m))$ before entering the non-linear fusion module; the detailed VP and fixed feature constructions are given in Appendix~\ref{sec:etas-feature-construction}. Observed $\Delta_i$ and $m_i$ are substituted only when the target-event likelihood terms are evaluated.

The Fusion model contains two non-linear branches: a temporal branch and a magnitude branch. In the temporal branch, the LSTM representation, the ETAS temporal feature and the candidate elapsed time are embedded and concatenated:

\begin{equation}
\boldsymbol{z}_{i}^{\mathrm{Fusion}}(\delta\mid H_i)
=
\mathrm{Concat}
\left[
\phi_h(\boldsymbol{h}_i),
\phi_t(I_i(\delta)),
\phi_{\Delta}(\delta)
\right],
\label{eq:fusion-temporal-feature}
\end{equation}

where $\phi_h$, $\phi_t$ and $\phi_{\Delta}$ denote embedding mappings. 
The concatenated temporal representation is then passed through the cumulative hazard function network (CHFN) to obtain the temporal cumulative hazard:

\begin{equation}
\Lambda_i^{\mathrm{Fusion}}(\delta\mid H_i)
=
\mathrm{CHFN}
\left(
\boldsymbol{z}_{i}^{\mathrm{Fusion}}(\delta\mid H_i)
\right),
\qquad \delta>0.
\label{eq:fusion-temporal-cumulative}
\end{equation}

This temporal branch allows Fusion to combine recent seismic history with ETAS-derived triggering information when modeling the temporal component of the forecasting process.

In the magnitude branch, Fusion does not directly introduce the LSTM hidden representation. Instead, the ETAS-derived magnitude feature, the candidate elapsed time and the candidate magnitude define the magnitude representation. The candidate magnitude is represented by its target-threshold exceedance $\tilde m=m-M_d$ and is then introduced as a differentiable input to the cumulative magnitude function network (CMFN). The magnitude representation is then constructed as

\begin{equation}
\boldsymbol{z}_{m,i}^{\mathrm{Fusion}}(\delta,m\mid H_i)
=
\mathrm{Concat}
\left[
\phi_m\!\left(G_i(m)\right),
\phi_{\Delta}(\delta),
\phi_M(\tilde m)
\right].
\label{eq:fusion-magnitude-feature}
\end{equation}

The corresponding cumulative magnitude distribution is

\begin{equation}
F_i^{\mathrm{Fusion}}(m\mid\delta,H_i)
=
\mathrm{CMFN}
\left(
\boldsymbol{z}_{m,i}^{\mathrm{Fusion}}(\delta,m\mid H_i)
\right).
\label{eq:fusion-magnitude-cumulative}
\end{equation}

The notation conditions both branches on the same forecasting history $H_i$, but the magnitude branch is architecturally constrained not to use the LSTM hidden state directly. This avoids transferring short-term sequential fluctuations into magnitude-density estimation, while still allowing non-linear correction of the Gutenberg--Richter prior.

Thus Fusion remains comparable with NPP while changing the information used to construct the cumulative functions: ETAS-derived triggering information augments the temporal branch, and Gutenberg--Richter-derived candidate-magnitude information informs the magnitude branch. The conversion from cumulative neural outputs to intensities, densities and likelihood scores is specified next.

\subsection{Target-event likelihood and evaluation metrics}
\label{subsec:target-events}
\label{subsec:evaluation-metrics}

The model architectures described above are constructed from the retained catalog determined by $M_{\mathrm{cut}}$. Following the retained-history and target-event forecasting protocol of ~\citet{ref44}, the target-event definition is not used to remove lower-magnitude retained events from the conditioning history. Instead, it specifies the point process whose likelihood is scored: lower-magnitude retained events remain in $H_i$, while only events above the forecasting threshold activate the event log-density terms. The forecasting target threshold is fixed at $M_d=3.0$ for all models and evaluations. Target events are identified by

\begin{equation}
\mathrm{mask}_i=\mathbb{I}(m_i\geq M_d),
\label{eq:target-mask}
\end{equation}

where $\mathbb{I}(\cdot)$ is the indicator function. Thus, $M_{\mathrm{cut}}$ determines the retained input history, whereas $M_d$ determines which retained events activate the event-occurrence and magnitude-density terms of the likelihood.

The target-event likelihood is the marked point-process likelihood of the process restricted to events with $m\geq M_d$. For a target event, the target-event marked conditional intensity is factorized into a temporal target-event intensity and a conditional target-event magnitude density:

\begin{equation}
\lambda_d^{\mathcal{M}}(t,m\mid H_t)
=
\lambda_{t,d}^{\mathcal{M}}(t\mid H_t)
\,
f_d^{\mathcal{M}}(m\mid t,H_t),
\qquad
m\in[M_d,M_{\max}],
\label{eq:rate}
\end{equation}

where $\mathcal{M}$ denotes ETAS, NPP or Fusion, and $H_t$ denotes the retained history available before the candidate event time $t$. The magnitude density is normalized over the target-event magnitude range:

\begin{equation}
\int_{M_d}^{M_{\max}}
f_d^{\mathcal{M}}(u\mid t,H_t)\,du
=
1 .
\label{eq:magnitude-normalization}
\end{equation}

The restriction from a retained-catalog marked process to Eq.~(\ref{eq:rate}), including the equivalent retained-interval likelihood form, is derived in Appendix~\ref{app:target-likelihood}. For ETAS, this restriction is closed form: the fitted temporal intensity is scaled by the Gutenberg--Richter probability mass above $M_d$, and the magnitude density is renormalized over $[M_d,M_{\max}]$ (Appendix~\ref{app:target-etas}). For NPP and Fusion, the target-event temporal hazard and magnitude distribution are learned directly through the likelihood below; the subscript $d$ denotes this target-restricted process.

For each retained interval $i$, the target-event cumulative temporal intensity over a candidate elapsed time $\delta$ is

\[
\Lambda_{d,i}^{\mathcal{M}}(\delta\mid H_i)
=
\int_{t_{i-1}}^{t_{i-1}+\delta}
\lambda_{t,d}^{\mathcal{M}}(s\mid H_s)\,ds .
\]

ETAS evaluates this quantity analytically, whereas the neural models use interval-indexed cumulative functions. From this subsection onward, the subscript $d$ denotes the fixed target threshold $M_d$. We write the target-event cumulative functions generically as $\Lambda_{d,i}^{\mathcal{M}}(\delta\mid H_i)$ and $F_{d,i}^{\mathcal{M}}(m\mid \delta,H_i)$, using the NPP and Fusion representations defined above. The temporal intensity and conditional magnitude density are obtained by differentiating with respect to candidate elapsed time and candidate magnitude, with observed values substituted only after differentiation:

\begin{eqnarray}
\lambda_{t,d}^{\mathcal{M}}(t_i\mid H_i)
&=&
\left.
\frac{\partial}{\partial \delta}
\Lambda_{d,i}^{\mathcal{M}}(\delta\mid H_i)
\right|_{\delta=\Delta_i},
\nonumber\\
f_d^{\mathcal{M}}(m_i\mid \Delta_i,H_i)
&=&
\left.
\frac{\partial}{\partial m}
F_{d,i}^{\mathcal{M}}(m\mid \Delta_i,H_i)
\right|_{m=m_i}.
\label{eq:neural-derivatives}
\end{eqnarray}

Using the retained-interval form of the target-event likelihood, the unified objective is

\begin{eqnarray}
\log L_{\mathcal{M}}
&=&
\sum_{i=1}^{N}
\Bigg[
\mathrm{mask}_i
\left(
\log \lambda_{t,d}^{\mathcal{M}}(t_i\mid H_i)
+
\log f_d^{\mathcal{M}}(m_i\mid \Delta_i,H_i)
\right)
-
\Lambda_{d,i}^{\mathcal{M}}(\Delta_i\mid H_i)
\Bigg].
\label{eq:target-ll}
\end{eqnarray}

Here $t_N$, the occurrence time of the final retained event in the test sequence, is also the endpoint of the evaluation window. The event log-intensity and magnitude log-density terms are included only for target events, whereas the cumulative temporal hazard is accumulated over all retained-event intervals through $t_N$ and is not multiplied by the target-event mask. Because the evaluation window ends at $t_N$, no additional post-event survival term is required. For reporting, we decompose Eq.~(\ref{eq:target-ll}) into temporal and magnitude components.

The mean temporal log-likelihood over target events is defined as

\begin{equation}
\mathrm{LL}_{\mathrm{temp}}^{\mathcal{M}}
=
\frac{1}{N_d}
\sum_{i=1}^{N}
\left[
\mathrm{mask}_i
\log \lambda_{t,d}^{\mathcal{M}}(t_i\mid H_i)
-
\Lambda_{d,i}^{\mathcal{M}}(\Delta_i\mid H_i)
\right],
\label{eq:ll-temp}
\end{equation}

where $N_d=\sum_{i=1}^{N}\mathrm{mask}_i$ is the number of target events in the test set. 
This metric rewards the model for assigning high temporal intensity to target events and penalizes excessive cumulative temporal intensity over the evaluated time intervals.

The mean magnitude log-likelihood is defined as

\begin{equation}
\mathrm{LL}_{\mathrm{mag}}^{\mathcal{M}}
=
\frac{1}{N_d}
\sum_{i=1}^{N}
\mathrm{mask}_i
\log f_d^{\mathcal{M}}(m_i\mid \Delta_i,H_i).
\label{eq:ll-mag}
\end{equation}

This metric evaluates how well the model represents the conditional magnitude distribution of target events. In ETAS, the magnitude density is determined by the Gutenberg--Richter law after target-event conditioning. In NPP and Fusion, the magnitude density is obtained by differentiating the neural cumulative magnitude function. The magnitude component is reported on the original likelihood scale, and higher values indicate a better conditional magnitude-density fit.

For the AVN temporal log-likelihood comparison, we also use a homogeneous Poisson benchmark with a constant target-event rate over the test period. Let $\mathrm{LL}_{\mathrm{temp}}^{\mathrm{Poi}}$ denote its mean temporal log-likelihood. The Poisson-relative temporal log-likelihood is

\begin{equation}
\Delta \mathrm{LL}_{\mathrm{temp}}^{\mathcal{M}}
=
\mathrm{LL}_{\mathrm{temp}}^{\mathcal{M}}
-
\mathrm{LL}_{\mathrm{temp}}^{\mathrm{Poi}}.
\label{eq:relative-temp-ll}
\end{equation}

A positive value indicates temporal forecasting skill beyond a constant-rate process. This relative score is used for Fig.~\ref{fig2:LL}; Fig.~\ref{fig5:benchmark_box} reports the original mean temporal log-likelihood $\mathrm{LL}_{\mathrm{temp}}^{\mathcal{M}}$ for direct benchmark comparison.

\subsection{Temporal information gain analysis}\label{subsec:tcig}

To further examine temporal forecasting differences between models, we use temporal information gain (TIG) and temporal cumulative information gain (TCIG). TIG compares the temporal log-likelihood contribution of two models at each target event, including the target-event log-intensity and the cumulative hazard over all retained intervals since the previous target event. TCIG accumulates these event-wise gains along the test sequence. The event-wise TIG distribution and the positive-event fraction are used to distinguish persistent cumulative improvement from gains concentrated in a small number of events. Full definitions are provided in Appendix~\ref{app:information-gain}.

\subsection{Experimental settings}\label{subsec:experimental-protocol}

All experiments use the target-event likelihood defined in Sect.~\ref{subsec:target-events}, with the forecasting threshold fixed at $M_d=3.0$. The input catalog cutoff $M_{\mathrm{cut}}$ is varied to change the retained historical sequence while keeping the target-event set unchanged. The compared models are ETAS, NPP and the proposed Fusion model. For a controlled comparison, NPP and Fusion use the same catalog partitions, retained input histories, fixed $M_{d}=3.0$ target-event sets, preprocessing procedure, target-event likelihood and evaluation metrics. The AVN experiments follow the forecasting protocol of ~\citet{ref44}, while the cross-catalog experiments follow the official EarthquakeNPP partitions~\citep{ref69}. ETAS parameter estimation, neural-input standardization and ETAS-feature standardization are performed using only the training portion of each catalog. During validation and testing, ETAS-derived features are evaluated causally from the ETAS feature history specified by the corresponding Fusion variant; the observed event time and magnitude enter only as likelihood-evaluation arguments.

For the AVN catalog, we evaluate two Fusion variants with the same neural architecture and likelihood. Fusion VP refits the ETAS parameters separately for each $M_{\mathrm{cut}}$ setting and computes ETAS-derived features from the corresponding current-cutoff history. Fusion fixed keeps the ETAS parameter set estimated at the reference threshold $M_{\mathrm{cut}}=3.0$ and computes ETAS-derived temporal features from the $m\geq3.0$ reference history, with a Gutenberg--Richter scale mapping to the current cutoff before standardization. This mapping expresses the raw ETAS feature on the current-cutoff scale; because it is a positive cutoff-specific constant and the normalization statistics are recomputed for each experiment, it does not change the standardized feature supplied to Fusion. Changing $M_{\mathrm{cut}}$ nevertheless changes the retained neural input catalog and the intervals on which the fixed ETAS feature is evaluated, while the ETAS parameters and feature history remain those of the reference $M_{\mathrm{cut}}=3.0$ setting. For the EarthquakeNPP benchmark datasets, we follow the official data partition: the Auxiliary and Training periods are used as training history, and the validation and testing periods follow the benchmark setting. Each dataset is evaluated at the official minimum cutoff, denoted min $M_{\mathrm{cut}}$, and at $M_{\mathrm{cut}}=3.0$. For each benchmark dataset and cutoff, neural models are trained on the retained catalog defined by that cutoff and evaluated on target events above $M_d=3.0$. In these benchmark experiments, a single ETAS parameter set estimated from the ComCat Auxiliary and Training periods at $M_{\mathrm{cut}}=3.0$ is used to generate the fixed ETAS prior features for all datasets and cutoff settings; the fixed temporal-feature history is also constructed at the $m\geq3.0$ reference cutoff for the catalog being evaluated.

For NPP and Fusion, results are reported as the mean and standard deviation over repeated random initializations unless otherwise specified. ETAS is deterministic after parameter estimation and is therefore reported without seed averaging. AVN temporal log-likelihoods are reported relative to the homogeneous Poisson benchmark, whereas EarthquakeNPP benchmark scores are reported on the original likelihood scale.

\section{Results}\label{sec2}
\subsection{AVN catalog and forecasting task}\label{secsub:AVN}

\begin{figure}[!htbp]
\centering
\includegraphics[width=0.95\textwidth]{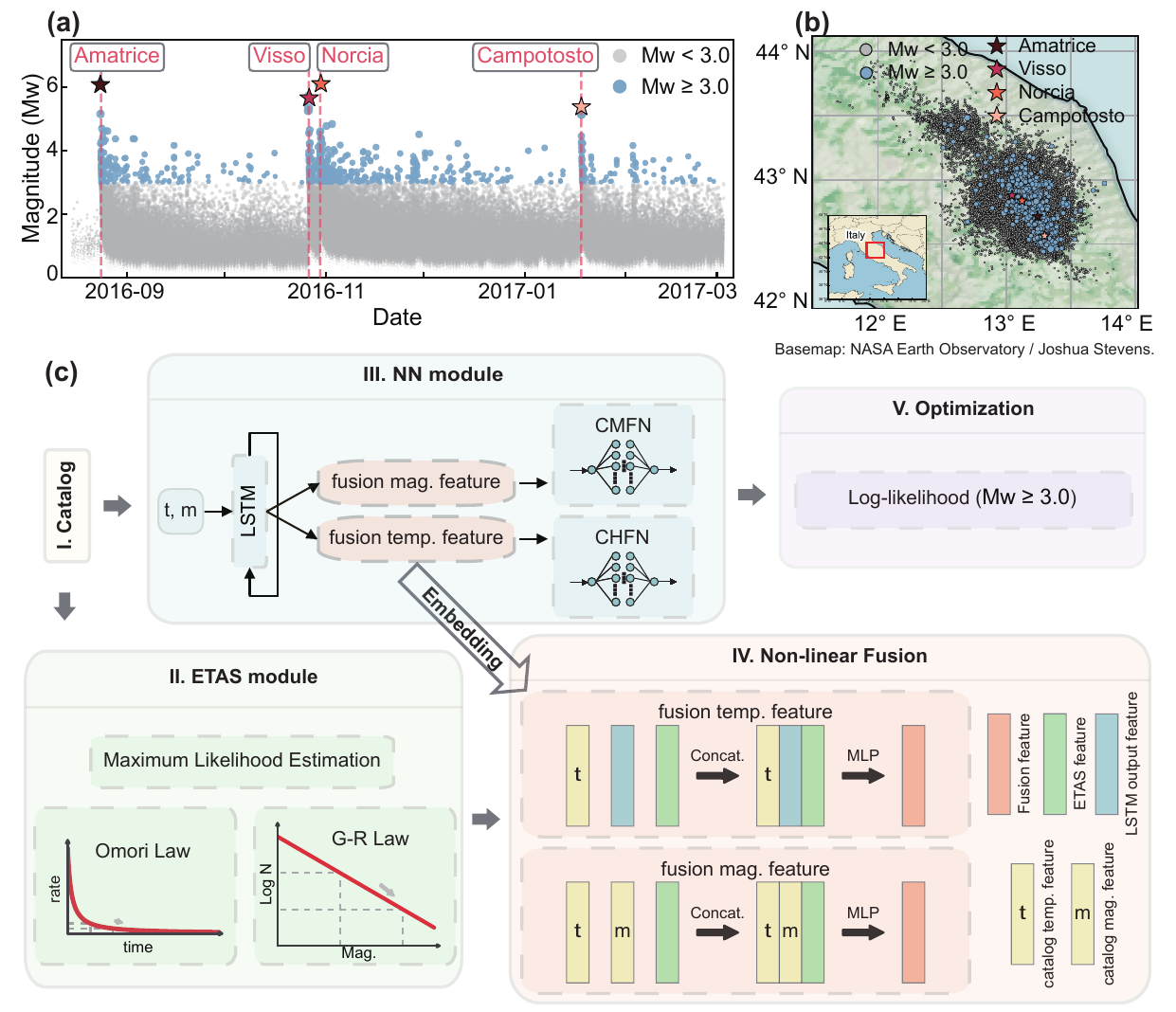}
\caption{\textbf{AVN earthquake sequence and scaling-law-informed Fusion framework.}
    \textbf{a}, Time--magnitude distribution of the 2016--2017 Amatrice--Visso--Norcia (AVN) earthquake sequence in the central Apennines, Italy. Vertical red dashed lines and stars mark the major seismic episodes near Amatrice, Visso, Norcia and Campotosto. Marker size is proportional to moment magnitude $M_w$. Blue markers denote target events with $M_w \geq 3.0$, whereas gray markers denote lower-magnitude events that can still enter the retained history when they satisfy the input cutoff $M_{\mathrm{cut}}$.
    \textbf{b}, Epicentral distribution and spatial extent of the AVN catalog used in this study. Gray and blue markers distinguish lower-magnitude events with \(M_w<3.0\) from target events with \(M_w\geq3.0\), respectively. The four major events are marked by stars. The inset map shows the location of the study region in central Italy. 
    \textbf{c}, Schematic of the proposed Fusion framework. Fusion is a neural marked point-process model informed by ETAS-derived scaling-law features: the temporal branch combines an LSTM history representation with an ETAS temporal-exposure feature summarizing Omori--Utsu triggering, while the magnitude branch uses a Gutenberg--Richter-informed candidate-magnitude feature without directly injecting the LSTM hidden state. The neural cumulative functions are differentiated to obtain the target-event temporal intensity and magnitude density used in likelihood evaluation.}
\label{fig1:framework}
\end{figure}

Figure~\ref{fig1:framework}a shows the 2016--2017 Amatrice--Visso--Norcia (AVN) earthquake sequence in the central Apennines, Italy~\citep{ref44}. The catalog contains several prominent seismic episodes, including the Amatrice $M_w$ 6.0 event on 24 August 2016, the Visso $M_w$ 5.9 event on 26 October 2016, the Norcia $M_w$ 6.5 event on 30 October 2016, and a sequence of moderate-to-large Campotosto events in early 2017. These events are followed by dense clusters of smaller earthquakes, producing the strong short-term triggering and aftershock relaxation pattern visible in the time--magnitude sequence.

Figure~\ref{fig1:framework}b shows the corresponding epicentral distribution within the central Apennines region, approximately spanning \(12^\circ\)--\(14^\circ\)~E and \(42^\circ\)--\(44^\circ\)~N. The four major events are marked separately, and gray and blue points distinguish lower-magnitude events from the \(M_w\geq3.0\) target-event set. The forecasting experiments are temporal marked point-process forecasts and do not use spatial coordinates as model inputs.

The enhanced catalog contains many lower-magnitude events whose completeness varies during the sequence, particularly around major earthquakes. Following the AVN forecasting design of ~\citet{ref44}, we therefore fix the likelihood-evaluation threshold at \(M_d=3.0\) and vary \(M_{\mathrm{cut}}\) only for the retained input history. This design tests whether lower-magnitude seismicity improves forecasts of a common target-event set; the preprocessing and target-event likelihood are described in Sect.~\ref{subsec:preprocess} and Sect.~\ref{subsec:target-events}.	
Figure~\ref{fig:S_AVN-log-log} compares post-event rate decay for the full catalog and the \(M_w\geq3.0\) target subset.

Figure~\ref{fig1:framework}c summarizes the Fusion architecture. Its temporal branch combines the recurrent history representation with an ETAS-derived triggering feature, whereas its magnitude branch uses a Gutenberg--Richter-informed candidate-magnitude feature without direct recurrent-history input. Fusion VP refits the ETAS component at each \(M_{\mathrm{cut}}\), while Fusion fixed retains the magnitude-complete \(M_{\mathrm{cut}}=3.0\) ETAS reference; both variants otherwise use the same neural architecture and target-event likelihood (Sect.~\ref{subsec:fusion} and Sect.~\ref{subsec:experimental-protocol}).

\subsection{AVN likelihood performance across input cutoffs}\label{secsub:LL}

\begin{figure}[!htbp]
\centering
\includegraphics[width=\textwidth]{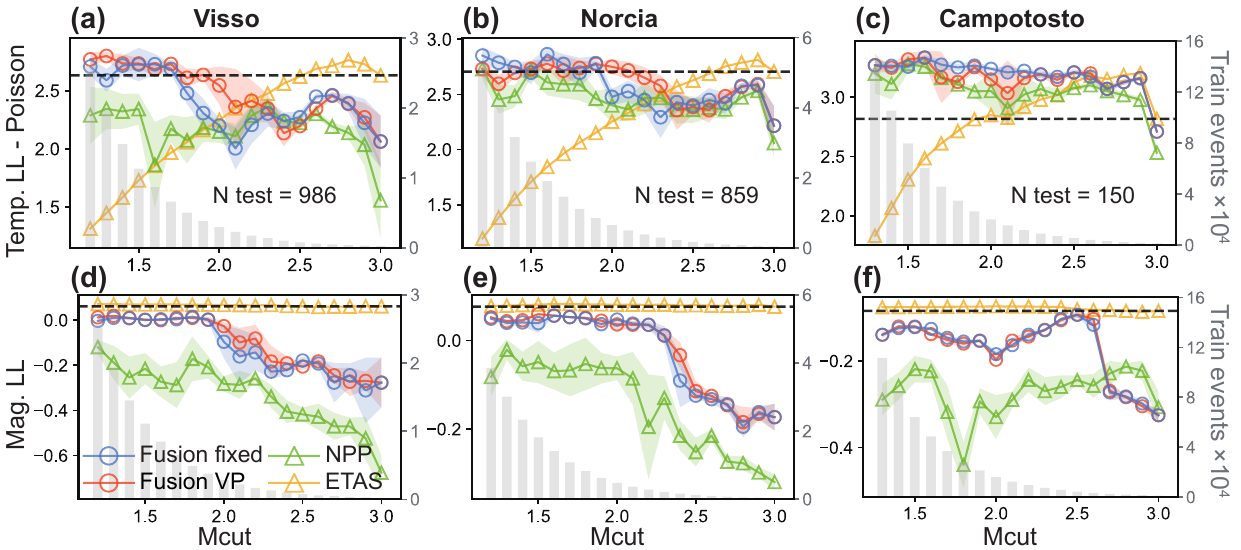}
\caption{\textbf{Target-event likelihood performance across input-history cutoffs on the AVN earthquake catalog.}
In all panels, the target-event threshold is fixed at $M_d=3.0$, and only the input-history cutoff $M_{\mathrm{cut}}$ is varied. Therefore, the same target events are evaluated at every $M_{\mathrm{cut}}$, while lower-magnitude events affect only the retained conditioning history. NPP denotes the neural point-process model of ~\citet{ref44}. 
\textbf{a--c}, Temporal log-likelihood relative to a homogeneous Poisson benchmark for the Visso, Norcia and Campotosto splits, respectively; positive values indicate better target-event timing forecasts than a constant-rate model.
\textbf{d--f}, Magnitude log-likelihood for the same splits, reported on the original likelihood scale.
All scores use the fixed target threshold $M_d=3.0$, while $M_{\mathrm{cut}}$ controls the retained input history. Fusion VP refits the ETAS component at each $M_{\mathrm{cut}}$; Fusion fixed uses the $M_{\mathrm{cut}}=3.0$ ETAS parameter set and reference feature history with a pre-standardization cutoff-scale mapping.
For NPP and Fusion, lines and markers show the mean over repeated random initializations and shaded bands show mean $\pm$ one standard deviation. ETAS is deterministic. Gray bars show the number of training events in units of $10^4$, annotated $N_{\mathrm{test}}$ values give the number of target events, and black dashed lines denote the fixed $M_{\mathrm{cut}}=3.0$ ETAS reference.}
\label{fig2:LL}
\end{figure}

Figure~\ref{fig2:LL} compares ETAS, NPP, Fusion VP and Fusion fixed on the three AVN test splits. Figures~\ref{fig2:LL}a--c report temporal log-likelihood relative to a homogeneous Poisson benchmark, whereas Figs.~\ref{fig2:LL}d--f report magnitude log-likelihood on the original scale. All methods are scored on the same \(M_d=3.0\) target events within each split, so variation with \(M_{\mathrm{cut}}\) reflects changes in the retained input history.

At low-to-intermediate \(M_{\mathrm{cut}}\), both Fusion variants generally achieve higher temporal likelihood than ETAS and NPP (Figs.~\ref{fig2:LL}a--c). Their advantage decreases as the retained catalog becomes sparser, and ETAS is competitive or higher in several settings above approximately \(M_{\mathrm{cut}}=2.3\). The timing advantage is therefore cutoff-dependent and is concentrated where lower-magnitude events are available as historical input.

For magnitude likelihood, Fusion is consistently higher than NPP and approaches the ETAS/Gutenberg--Richter reference at the lowest cutoffs, but ETAS remains the strongest or near-strongest model overall (Figs.~\ref{fig2:LL}d--f). At \(M_{\mathrm{cut}}=1.2\) in the Visso split, for example, the mean Fusion magnitude likelihood remains approximately 0.05 below ETAS. These comparisons provide no evidence that catalog history adds stable predictive information about target-event magnitude beyond the Gutenberg--Richter distribution; the difference between Fusion and NPP instead reflects a reduction in the purely neural model's magnitude-density deficit relative to that reference. Consistently, the Fusion-m ablation, which also supplies recurrent history to the magnitude branch, gives generally weaker and more variable magnitude scores (Fig.~\ref{fig:S_LL}).

Fusion fixed and Fusion VP have similar mean performance across most cutoffs, and Fusion fixed has narrower variability in several settings. Their comparable results support the use of a magnitude-complete reference ETAS calibration across input cutoffs. Results for Campotosto are less precise because its test split contains only $N_{\mathrm{test}}=150$ target events.

\subsection{Temporal gains}\label{secsub:TCIG}

Temporal cumulative information gain (TCIG) tracks the accumulated event-wise temporal log-likelihood difference between Fusion and a baseline along each test sequence (Sect.~\ref{subsec:tcig} and Appendix~\ref{app:information-gain}). Positive final values indicate a net temporal likelihood gain for Fusion.

\begin{figure}[!htb]
\centering
\includegraphics[width=\textwidth]{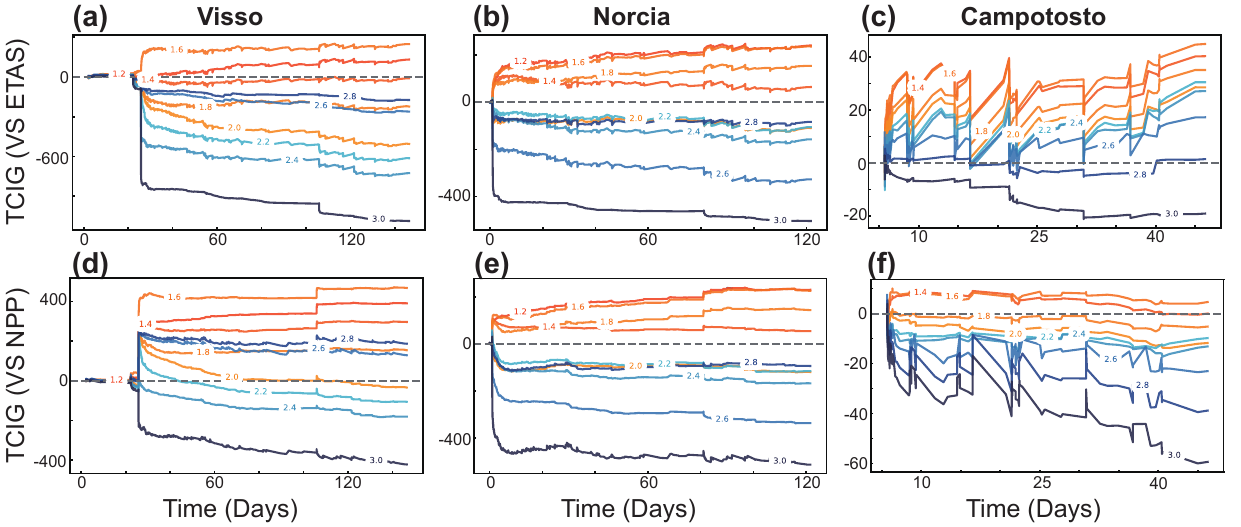}
\caption{\textbf{Temporal cumulative information gain of Fusion on the AVN earthquake sequence.}
\textbf{a--c}, Temporal cumulative information gain (TCIG) of Fusion relative to ETAS for the Visso, Norcia and Campotosto splits, respectively, with ETAS shown under the fixed $M_{\mathrm{cut}}=3.0$ reference setting.
\textbf{d--f}, TCIG of Fusion relative to NPP for the same splits, with NPP evaluated at $M_{\mathrm{cut}}=1.2$ for Visso and Norcia and $M_{\mathrm{cut}}=1.3$ for Campotosto.
TCIG accumulates the event-wise temporal log-likelihood difference along target-event time, including both target-event log-intensity and the cumulative target-event temporal intensity over all retained intervals between successive target events.
The horizontal axis gives elapsed time in days after the first target event in the test set.
Different colors denote Fusion models trained with different retained-catalog cutoffs $M_{\mathrm{cut}}$.
The gray dashed line marks TCIG$=0$; values above zero indicate cumulative temporal likelihood gain of Fusion over the corresponding baseline.}
\label{fig3:TCIG}
\end{figure}

Relative to ETAS, Fusion accumulates positive temporal information gain at several low-to-intermediate $M_{\mathrm{cut}}$ settings, most clearly in the Visso and Norcia splits (Figs.~\ref{fig3:TCIG}a--c). The high-cutoff curves are weaker or negative, consistent with the cutoff dependence in Fig.~\ref{fig2:LL}. Relative to NPP, Fusion also accumulates positive gain over broad portions of the Visso and Norcia sequences (Figs.~\ref{fig3:TCIG}d, e).	
Figure~\ref{fig:S_temp_rate} shows the corresponding temporal intensities: Fusion maintains elevated intensity across several aftershock-active intervals, whereas NPP exhibits more isolated peaks.

The Campotosto curves fluctuate more strongly and show positive or near-zero final gains only for some lower-cutoff settings (Figs.~\ref{fig3:TCIG}c, f). Its smaller target-event sample makes the cumulative curves more sensitive to individual earthquakes. Overall, the TCIG curves show that the mean temporal gains in Fig.~\ref{fig2:LL} often accumulate across successive target events rather than arising from a single isolated event.

\begin{figure}[!htbp]
\centering
\includegraphics[width=\textwidth]{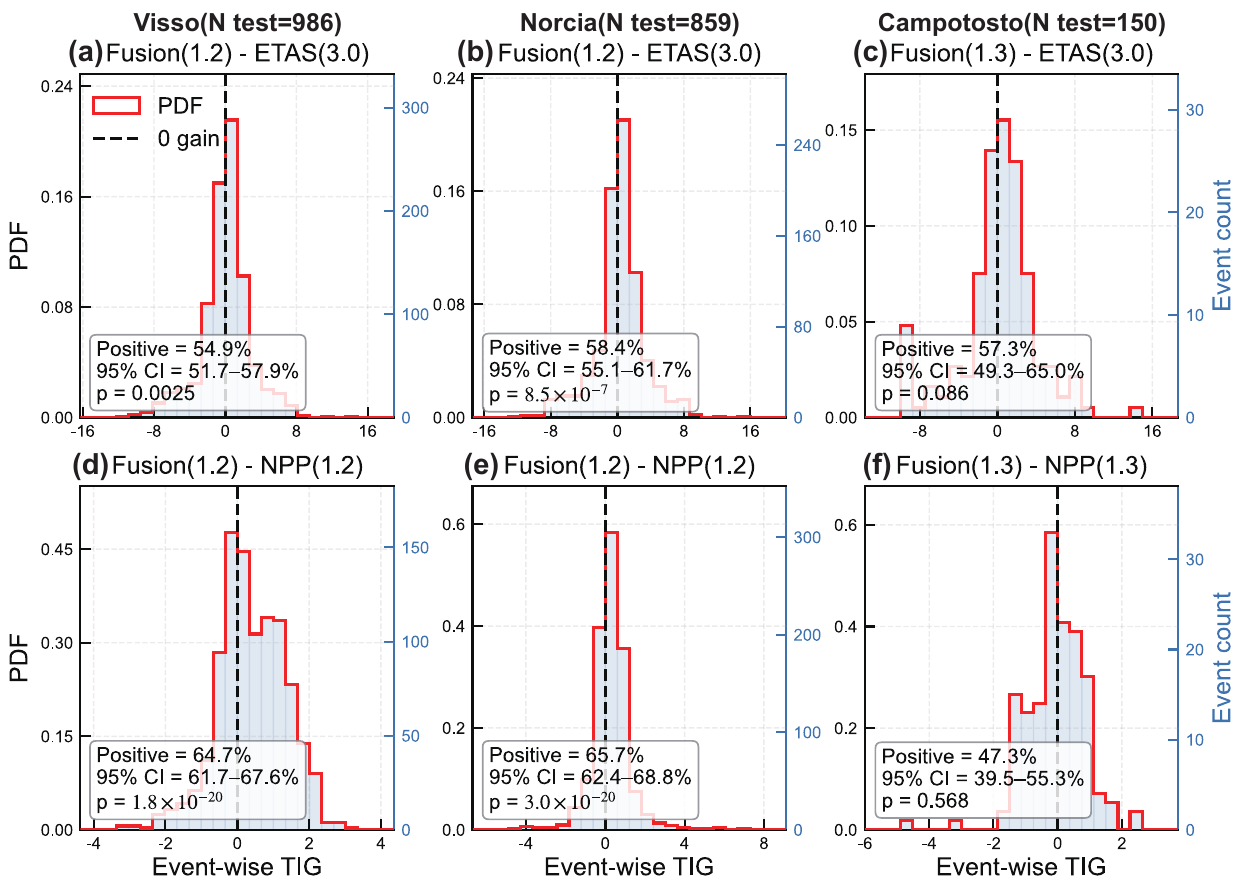}
\caption{\textbf{Event-wise temporal information gain distributions on AVN target events.}
Each histogram shows the temporal information gain (TIG) for individual test target events in the Visso, Norcia and Campotosto splits.
TIG is defined as the event-wise temporal log-likelihood difference between the first and second model in a comparison, after accounting for the cumulative target-event temporal intensity over all retained intervals since the previous target event.
\textbf{a--c}, Fusion--ETAS comparisons for the Visso, Norcia and Campotosto splits, respectively. Fusion is evaluated at the low-cutoff settings, \(M_{\mathrm{cut}}=1.2\) for Visso and Norcia and \(M_{\mathrm{cut}}=1.3\) for Campotosto, whereas ETAS is evaluated at the \(M_{\mathrm{cut}}=3.0\) reference setting.
\textbf{d--f}, Fusion--NPP comparisons for the same splits, with both neural models evaluated at the corresponding low-cutoff settings.
The dashed vertical line marks zero gain. Positive TIG values indicate target events for which the first model assigns higher temporal likelihood than the second model.
Red outlines show the estimated TIG probability density, and the light blue bars show the corresponding event counts.
The annotation in each panel reports the percentage of events with TIG \(>0\), the corresponding 95\% confidence interval, and the nominal two-sided sign-test \(p\) value.
The calculation method is given in Appendix~\ref{app:information-gain}.}
\label{fig4:tig_PDF}
\end{figure}

Figure~\ref{fig4:tig_PDF} decomposes the selected-model comparisons into event-wise temporal information gain (TIG). Fusion and NPP use the low-cutoff settings with strong temporal likelihood performance, while ETAS uses the magnitude-complete $M_{\mathrm{cut}}=3.0$ reference. The distributions therefore describe these selected configurations rather than all cutoff settings. A positive TIG means that the first model assigns higher temporal likelihood to that target event (Sect.~\ref{subsec:tcig} and Appendix~\ref{app:information-gain}).

\noindent\textbf{Event-wise gains.}

Figure~\ref{fig4:tig_PDF} further decomposes the temporal forecasting differences into event-wise temporal information gain (TIG) distributions. 
The figure is computed at selected input-cutoff settings where the corresponding models show strong temporal likelihood performance: Fusion and NPP use the lowest retained-history cutoffs, \(M_{\mathrm{cut}}=1.2\) for the Visso and Norcia splits and \(M_{\mathrm{cut}}=1.3\) for the Campotosto split, whereas ETAS is evaluated using the conservative high-cutoff reference \(M_{\mathrm{cut}}=3.0\). Thus, Fig.~\ref{fig4:tig_PDF} compares the models under their favorable or reference cutoff configurations rather than summarizing TIG distributions over all \(M_{\mathrm{cut}}\) values.
TIG is the per-target-event version of TCIG: a positive value means that the first model in the comparison assigns higher temporal likelihood to that target event, after accounting for the cumulative target-event temporal intensity over the retained intervals since the previous target event. This analysis asks whether Fusion wins broadly across target events, or whether its average advantage is produced by a small number of large gains. The TIG definitions are provided in Sect.~\ref{subsec:tcig} and Appendix~\ref{app:information-gain}.

Fusion assigns higher temporal likelihood than ETAS to 54.9\%, 58.4\% and 57.3\% of the Visso, Norcia and Campotosto target events, respectively (Figs.~\ref{fig4:tig_PDF}a--c). Relative to NPP, the corresponding positive-event fractions are 64.7\%, 65.7\% and 47.3\% (Figs.~\ref{fig4:tig_PDF}d--f). Thus, the Fusion advantage is distributed across a majority of events in Visso and Norcia, while the Campotosto Fusion--NPP comparison remains uncertain.

The nominal two-sided sign-test $p$ values are 0.0025, $8.5\times10^{-7}$ and 0.086 for Fusion--ETAS, and $1.8\times10^{-20}$, $3.0\times10^{-20}$ and 0.568 for Fusion--NPP, in split order (Fig.~\ref{fig4:tig_PDF}; Table~\ref{tab:TIG-value}). Because earthquake events are temporally correlated and the plotted cutoffs were selected for strong likelihood performance, these values are descriptive diagnostics rather than formal independent-event significance tests. The Campotosto fractions are especially sensitive to sampling variability because this split contains only 150 target events.

Figure~\ref{fig:S_tig_PDF_origin} provides the direct NPP--ETAS comparison. Its positive-event fractions are 43.6\% for Visso, 50.8\% for Norcia and 59.3\% for Campotosto, showing greater split-to-split variation than the Fusion--ETAS comparison. Together with Fig.~\ref{fig4:tig_PDF}, this result is consistent with the ETAS-derived feature contributing to the event-wise temporal gains beyond the use of a neural point-process architecture alone.

\subsection{Performance across benchmark catalogs}\label{secsub:benchmark}

\begin{figure}[!htb]
\centering
\includegraphics[width=\textwidth]{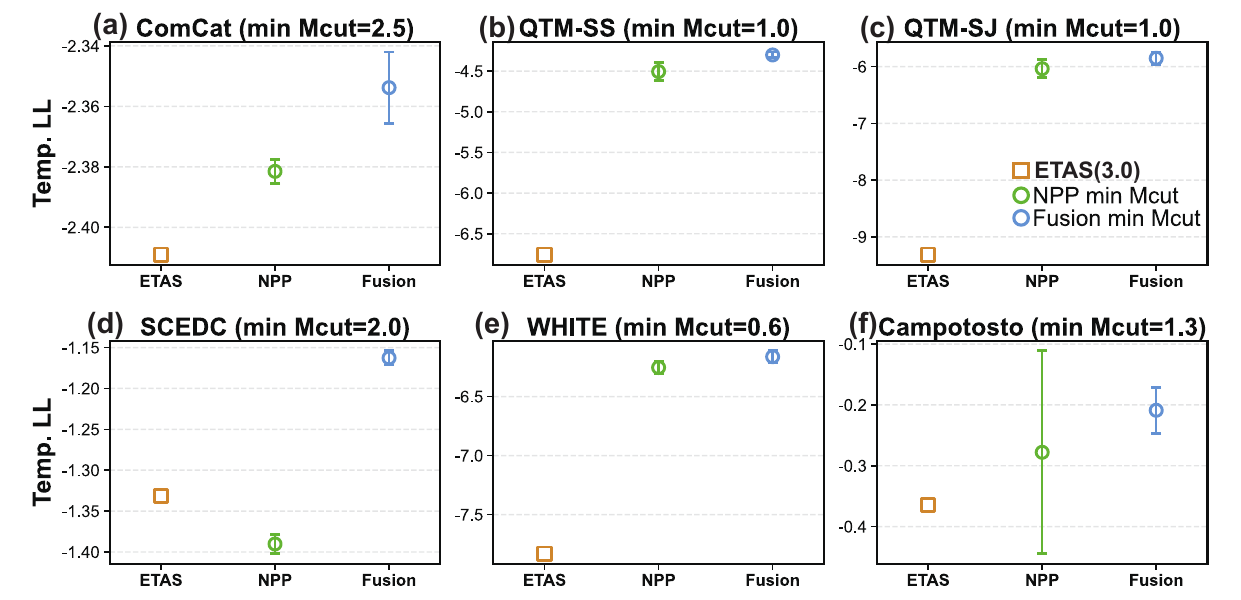}
\caption{
\textbf{Target-event temporal log-likelihood comparison across the EarthquakeNPP benchmark catalogs and the Campotosto AVN split.}
Panels \textbf{a--f} correspond to ComCat, QTM-SaltonSea (QTM-SS), QTM-SanJacinto (QTM-SJ), SCEDC, WHITE and Campotosto, respectively. NPP and Fusion use the dataset-specific minimum \(M_{\mathrm{cut}}\) indicated above each panel. ETAS uses the magnitude-complete \(M_{\mathrm{cut}}=3.0\) reference configuration, which provides its strongest reliable result under the catalog-completeness requirement. Markers show the mean over repeated random initializations and error bars show mean \(\pm\) one standard deviation for NPP and Fusion; ETAS is deterministic. For the five California catalogs, Fusion uses the shared fixed ETAS prior estimated from the ComCat training history at \(M_{\mathrm{cut}}=3.0\); Campotosto uses the corresponding AVN fixed-prior configuration. All temporal likelihoods are evaluated on target events with $m\geq M_d=3.0$. Higher temporal log-likelihood (Temp.\ LL) indicates better target-event timing performance.}
\label{fig5:benchmark_box}
\end{figure}

We next evaluated the models on five EarthquakeNPP benchmark catalogs: ComCat, QTM-SaltonSea, QTM-SanJac, SCEDC and WHITE~\citep{ref69}. Figure~\ref{fig5:benchmark_box} compares their target-event temporal likelihoods with the Campotosto AVN split, which is included because it has the longest training history among the three AVN splits. The catalogs span different spatial domains, durations, event counts and minimum input cutoffs (Fig.~\ref{fig:S_benchmark_spatial} and Table~\ref{tab:benchmark-spatial-domain}).

The comparison uses model-specific input configurations. NPP and Fusion use the benchmark minimum cutoffs to retain lower-magnitude history, whereas ETAS uses $M_{\mathrm{cut}}=3.0$, its strongest reliable configuration because ETAS calibration depends on magnitude completeness. Fusion uses one ETAS parameter set estimated from the ComCat training catalog for all five California datasets; details of the partitions and fixed-prior protocol are given in Sect.~\ref{subsec:preprocess} and Sect.~\ref{subsec:experimental-protocol}.

Under these configurations, Fusion has the highest mean temporal log-likelihood in all five benchmark catalogs and in Campotosto (Fig.~\ref{fig5:benchmark_box}). This pattern is observed despite the shared ComCat ETAS parameter set and the heterogeneity of the regional catalogs. Together with the cutoff-dependent AVN results, it shows that Fusion's timing advantage is concentrated in configurations that retain lower-magnitude historical events.

The magnitude comparison shows a different pattern. Fusion has higher magnitude log-likelihood than NPP in all five benchmark catalogs and Campotosto, whereas the ETAS/Gutenberg--Richter reference is highest in five of the six panels; Fusion is highest only for ComCat (Fig.~\ref{fig:S_benchmark_box_mag} and Appendix~\ref{app:magnitude-results}). The cross-catalog results therefore provide no consistent evidence of predictive information about target-event magnitude beyond the Gutenberg--Richter distribution. Fusion's advantage over NPP primarily narrows the deficit of the purely neural magnitude density relative to this reference. Further comparisons related to Fig.~\ref{fig5:benchmark_box} are provided in Table~\ref{tab:benchmark-likelihood-results}, which reports the temporal and magnitude log-likelihoods of Fusion, NPP, and ETAS for the five EarthquakeNPP benchmark catalogs and the representative Campotosto AVN split under both the benchmark-specific minimum input cutoff and the common maximum cutoff \(M_{\mathrm{cut}}=3.0\). For Campotosto, the reported temporal and magnitude results correspond to Figs.~\ref{fig2:LL}c and ~\ref{fig2:LL}f, respectively.
 
\section{Discussion}
The principal result is that empirical seismicity laws and neural sequence learning provide complementary information for target-event timing. On the Amatrice--Visso--Norcia sequence, Fusion gives higher temporal likelihood than both ETAS and NPP when low-to-intermediate input cutoffs retain small-to-moderate earthquakes. Its advantage decreases as the retained history becomes sparse, and ETAS remains competitive at higher cutoffs. Across the benchmark catalogs, Fusion also gives the highest mean temporal likelihood under the model-specific configurations evaluated. By contrast, the magnitude comparisons provide no consistent evidence of predictive information beyond the Gutenberg--Richter distribution.

The cutoff dependence clarifies the contribution of lower-magnitude seismicity. Holding $M_d=3.0$ fixed ensures that the same target events are scored as $M_{\mathrm{cut}}$ changes, so the observed temporal differences arise from the historical information available to the models rather than from a changing forecast target. Lower-magnitude events provide a denser description of sequence evolution between target events and can carry information about aftershock relaxation and short-term triggering. The cumulative and event-wise diagnostics show that the Fusion gains are distributed through substantial portions of the Visso and Norcia sequences, which is consistent with this historical information contributing repeatedly rather than through a few isolated events. These diagnostics do not identify a unique physical mechanism, but they support an interpretation in which the recurrent representation extracts additional sequence dependence while the ETAS-derived feature supplies an empirical triggering reference. At higher $M_{\mathrm{cut}}$, the retained history becomes sparser and provides less information for estimating such departures, leaving the parametric ETAS model more competitive.

The contrast between temporal and magnitude likelihoods identifies where the additional forecasting information lies. Target-event timing can depend on the recent retained history, making the combination of recurrent history and ETAS-derived temporal information useful. Target-event magnitudes remain well described by the Gutenberg--Richter form. Although Fusion reduces the magnitude-likelihood deficit of NPP, the ETAS/Gutenberg--Richter reference remains the stronger model overall, and the weaker and more variable Fusion-m ablation shows that directly adding recurrent history does not yield consistent magnitude gains. The magnitude branch therefore completes the marked point-process likelihood and tests for departures from Gutenberg--Richter scaling, but it does not demonstrate additional conditional magnitude predictability in these experiments. The comparable temporal performance of Fusion fixed and Fusion VP further suggests that a magnitude-complete reference ETAS calibration can serve as a stable temporal-feature source across input cutoffs. The use of a shared ComCat ETAS parameter set in the California benchmarks extends this observation across heterogeneous catalogs, although it demonstrates consistency under the tested protocol rather than complete regional transferability.

\conclusions
\label{sec:conclusions}
These results support using empirical seismicity laws as structured inputs to neural point processes for target-event timing. The timing advantage depends on an informative lower-magnitude history, whereas the magnitude results show no detectable incremental information beyond the Gutenberg--Richter distribution. These conclusions are based on retrospective, nonspatial forecasts and model-specific benchmark configurations, and the value of lower-magnitude events may depend on catalog completeness. Pseudo-prospective evaluation with explicit treatment of catalog uncertainty and spatially marked forecasts is needed to assess the operational value of this scaling-law-informed strategy.



\clearpage
\appendix









\section{Mainshock-referenced Omori-type rate decay in the AVN sequence} 

\begin{figure}[!htb]
\centering
\includegraphics[width=\textwidth]{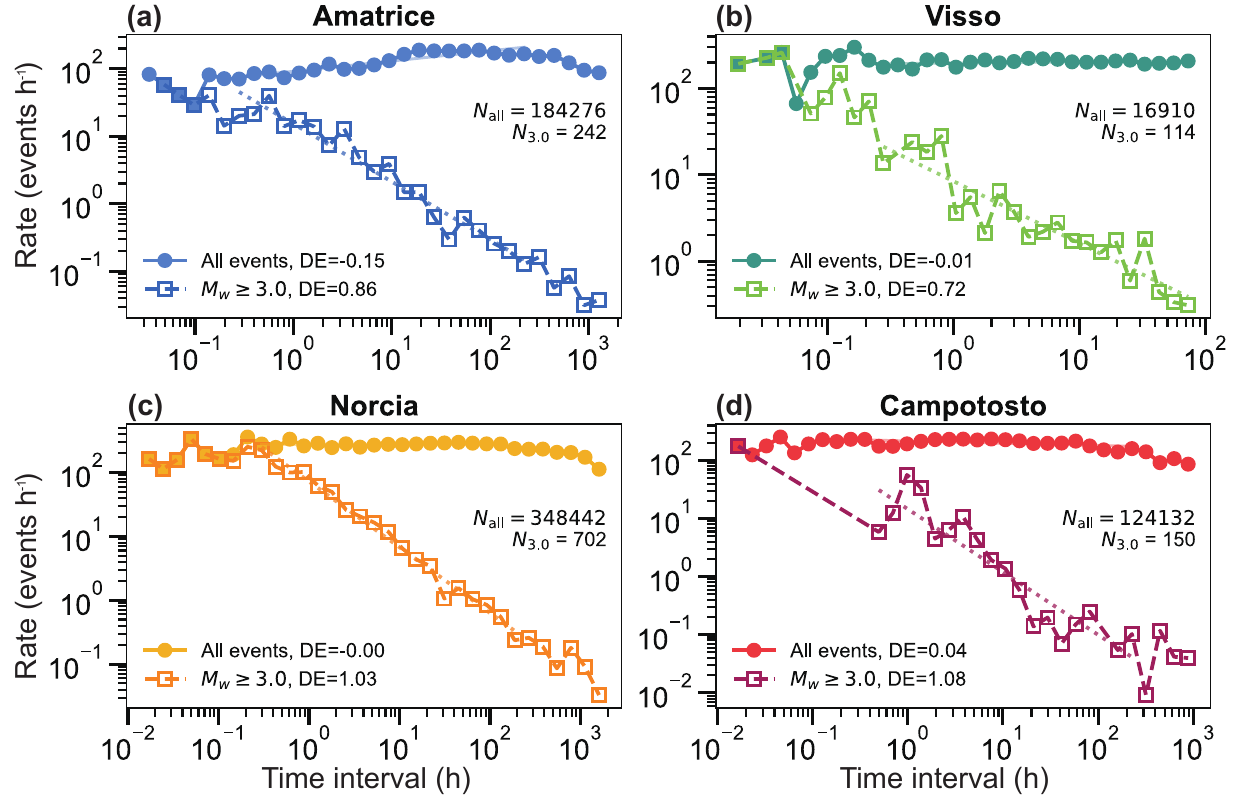}
\caption{\textbf{Mainshock-referenced Omori-type rate decay in the AVN sequence.} Log--log plots of seismicity rate as a function of elapsed time from each major AVN event, with time measured in hours.
\textbf{a--d}, Results for the Amatrice, Visso, Norcia, and Campotosto episodes, respectively. Filled circles show rates calculated from all retained catalog events, whereas open squares show rates calculated from target events with \(M_w\geq3.0\). The rate in each time bin is defined as the number of events within the bin divided by the bin duration in hours. \(N_{\mathrm{all}}\) and \(N_{3.0}\) denote the numbers of all retained events and \(M_w\geq3.0\) target events included in each analysis window, respectively. DE denotes the fitted decay exponent obtained from the log--log rate--time relation.}
\label{fig:S_AVN-log-log}
\end{figure}

To further characterize the temporal structure of the AVN catalog shown in main text Fig.~\ref{fig1:framework}a,b, we examine the seismicity-rate decay following the major AVN seismic episodes. Figure~\ref{fig:S_AVN-log-log} shows log--log plots of event rate as a function of elapsed time from each corresponding major event, measured in hours. Panels \textbf{a}--\textbf{d} follow the chronological order of the four major episodes: Amatrice, Visso, Norcia and Campotosto. For each panel, the analysis window starts from the corresponding major event and extends to the next major event, except for Campotosto, which extends to the end of the catalog.

For each time bin, the seismicity rate is calculated as the number of events within the bin divided by the bin duration in hours and is therefore reported in events per hour. Rates are calculated separately for all retained catalog events and for target events satisfying \(M_w\geq3.0\). The corresponding event counts are reported in each panel as \(N_{\mathrm{all}}\) and \(N_{3.0}\), respectively. These counts illustrate the substantial difference in sample size between the full retained catalog and the target-event subset.

The decay exponent (DE) is estimated from the linear relation between the logarithm of the seismicity rate and the logarithm of elapsed time. Under the adopted sign convention, a larger positive DE indicates a stronger decrease in seismicity rate with time. For the \(M_w\geq3.0\) target events, the estimated DE values are 0.86, 0.72, 1.03, and 1.08 for the Amatrice, Visso, Norcia, and Campotosto windows, respectively. The consistently positive values indicate a clearer Omori-type rate decay in the target-event sequence.

By contrast, the DE values obtained from all retained events are close to zero and range from slightly negative to slightly positive, indicating a substantially flatter rate--time relation. This contrast may partly reflect the dominance of lower-magnitude events in the enhanced catalog and the greater sensitivity of these events to short-term variations in catalog completeness following large earthquakes.

This analysis provides additional support for using \(M_d=3.0\) as a conservative target-event threshold in the likelihood evaluation. Events with \(M_w\geq3.0\) exhibit a more clearly resolved aftershock-rate decay, whereas lower-magnitude events remain available as additional historical information in the retained input sequence.

\section{ETAS feature construction for Fusion VP and Fusion fixed}
\label{sec:etas-feature-construction}

This note defines the ETAS-derived scalar features used by the Fusion architecture in the ``Scaling-law-informed Fusion model'' subsection of Methods. The construction follows the model-building order in the main text. The current input cutoff $M_{\mathrm{cut}}$ defines the retained catalog used by the neural history encoder. The ETAS feature history depends on the Fusion variant:
\[
H_t^{\mathrm{cut}}
=
\{(t_j,m_j):t_j<t,\ m_j\geq M_{\mathrm{cut}}\},
\qquad
H_t^{3.0}
=
\{(t_j,m_j):t_j<t,\ m_j\geq 3.0\}.
\]
Fusion VP uses the current-cutoff ETAS feature history $H_t^{\mathrm{cut}}$, whereas Fusion fixed uses the reference ETAS feature history $H_t^{3.0}$. When the current cutoff is lower than 3.0, events with $M_{\mathrm{cut}}\leq m_j<3.0$ remain part of the retained catalog used by the LSTM history representation. The temporal ETAS feature in Fusion fixed is built from $H_t^{3.0}$. Before standardization, Eq.~(\ref{eq:fusion-fixed-scale-alignment}) expresses the reference intensity on the current-cutoff scale while preserving the temporal pattern implied by the $m\geq3.0$ ETAS form. The target-event threshold $M_d$ is introduced in Sect.~\ref{subsec:target-events} and Appendix~\ref{app:target-likelihood}; it is not needed to define the retained-catalog ETAS features themselves.

For a candidate time $t=t_{i-1}+\delta$ in retained interval $i$, let $\lambda_0^{\star}(t\mid H_t^{\star})$ denote the ETAS temporal intensity used to build the Fusion feature, where $\star\in\{\mathrm{VP},\mathrm{fixed}\}$, $H_t^{\mathrm{VP}}=H_t^{\mathrm{cut}}$ and $H_t^{\mathrm{fixed}}=H_t^{3.0}$. The subscript 0 indicates that this quantity is on the retained-catalog scale before target-event scoring. The corresponding candidate-interval temporal integral is

\begin{equation}
\Lambda_i^{\star}(\delta\mid H_i^{\star})
=
\int_{t_{i-1}}^{t_{i-1}+\delta}
\lambda_0^{\star}(s\mid H_s^{\star})
\,ds .
\label{eq:fusion-etas-temporal-integral}
\end{equation}

This integral is defined for an arbitrary candidate elapsed time \(\delta\), conditional on the history available before the candidate time. During likelihood evaluation, the observed elapsed time \(\Delta_i\) is substituted only as the candidate argument. The scalar temporal feature supplied to the Fusion temporal branch is standardized as

\begin{equation}
I_i^{\star}
\left(
\delta\mid H_i^{\star}
\right)
=
\frac{
\Lambda_i^{\star}
\left(
\delta\mid H_i^{\star}
\right)
-
\mu_{\mathrm{et}}^{\star}
}{
\sigma_{\mathrm{et}}^{\star}
},
\\[4pt]
\phi_t
\left[
I_i^{\star}
\left(
\delta\mid H_i^{\star}
\right)
\right]
\in
\mathbb{R}^{64}.
\label{eq:fusion-etas-time-feature}
\end{equation}

where $\mu_{\mathrm{et}}^{\star}$ and $\sigma_{\mathrm{et}}^{\star}$ are computed from the training portion of the corresponding experiment.

The magnitude feature is constructed from a Gutenberg--Richter cumulative quantity on the current input-cutoff scale:
\begin{equation}
G_i^{\star}(m)
=
\frac{
F_i^{\star}(m)-\mu_{\mathrm{em}}^{\star}
}{
\sigma_{\mathrm{em}}^{\star}
},
\qquad
\phi_m(G_i^{\star}(m))\in\mathbb{R}^{64},
\label{eq:fusion-etas-mag-feature}
\end{equation}
where $m$ is a candidate magnitude and $\mu_{\mathrm{em}}^{\star}$ and $\sigma_{\mathrm{em}}^{\star}$ are training-set normalization constants.

In Fusion VP, the ETAS parameter set is refitted for each current input cutoff and is denoted by $\theta_{M_{\mathrm{cut}}}$. The temporal intensity used for the feature is therefore
\begin{equation}
\lambda_0^{\mathrm{VP}}(t\mid H_t^{\mathrm{cut}})
=
\lambda_0^{\mathrm{ETAS}}
\left(t\mid H_t^{\mathrm{cut}};\theta_{M_{\mathrm{cut}}}\right),
\label{eq:fusion-vp-retained-intensity}
\end{equation}
and the temporal scalar in Eq.~(\ref{eq:fusion-etas-time-feature}) is obtained by substituting Eq.~(\ref{eq:fusion-vp-retained-intensity}) into Eq.~(\ref{eq:fusion-etas-temporal-integral}). The VP magnitude quantity uses the Gutenberg--Richter parameter estimated at the same current cutoff:
\begin{equation}
F_i^{\mathrm{VP}}(m)
=
\frac{
1-\exp[-\beta_{M_{\mathrm{cut}}}(m-M_{\mathrm{cut}})]
}{
1-\exp[-\beta_{M_{\mathrm{cut}}}(M_{\max}-M_{\mathrm{cut}})]
},
\qquad
M_{\mathrm{cut}}\leq m\leq M_{\max}.
\label{eq:fusion-vp-mag-feature}
\end{equation}

In Fusion fixed, the ETAS parameter set is estimated once at the reference cutoff $M_{\mathrm{cut}}=3.0$ and is kept fixed as $\theta_{3.0}$. The ETAS feature history is also formed at the reference cutoff, $H_t^{3.0}$. The fixed-parameter temporal intensity
\[
\lambda_{3.0}^{\mathrm{ETAS}}(t\mid H_t^{3.0};\theta_{3.0})
\]
is on the $m\geq3.0$ reference scale. When the current input cutoff is lower than 3.0, the corresponding current-cutoff intensity implied by the same Gutenberg--Richter scaling is obtained using
\begin{equation}
q_{3.0}(M_{\mathrm{cut}};\beta_{3.0}) \\
=
\Pr(M\geq3.0\mid M\geq M_{\mathrm{cut}})
\\
=
\frac{
\exp[-\beta_{3.0}(3.0-M_{\mathrm{cut}})]
-
\exp[-\beta_{3.0}(M_{\max}-M_{\mathrm{cut}})]
}{
1-\exp[-\beta_{3.0}(M_{\max}-M_{\mathrm{cut}})]
}.
\label{eq:fusion-fixed-threshold-factor}
\end{equation}
The cutoff-aligned fixed temporal intensity is then
\begin{equation}
\lambda_0^{\mathrm{fixed}}(t\mid H_t^{3.0})
=
\frac{
\lambda_{3.0}^{\mathrm{ETAS}}(t\mid H_t^{3.0};\theta_{3.0})
}{
q_{3.0}(M_{\mathrm{cut}};\beta_{3.0})
}.
\label{eq:fusion-fixed-scale-alignment}
\end{equation}
The factor $q_{3.0}$ is a pre-standardization cutoff-scale mapping for the Fusion fixed prior feature. Although it has the same Gutenberg--Richter threshold-mass form as the target-conditioning probability $p_d$ in Eq.~(\ref{eq:pd}) when $M_d=3.0$, its role is different: $q_{3.0}$ expresses the reference $m\geq3.0$ ETAS temporal feature on the current input-cutoff scale, whereas $p_d$ restricts the likelihood to the fixed target-event process.

Because $q_{3.0}$ is a positive constant within each cutoff experiment, it cancels when the mapped temporal integrals are standardized using statistics from that experiment. Writing $\Lambda_{3.0,i}$ for the unmapped reference integral and $\bar\mu_{3.0}$ and $\bar\sigma_{3.0}$ for its mean and standard deviation over the same training intervals gives
\[
\frac{\Lambda_{3.0,i}/q_{3.0}-\bar\mu_{3.0}/q_{3.0}}
{\bar\sigma_{3.0}/q_{3.0}}
=
\frac{\Lambda_{3.0,i}-\bar\mu_{3.0}}{\bar\sigma_{3.0}}.
\]
Thus $q_{3.0}$ does not alter the standardized scalar supplied to Fusion. It is retained to make the theoretical cutoff mapping of the unstandardized ETAS intensity explicit. Substituting Eq.~(\ref{eq:fusion-fixed-scale-alignment}) into Eq.~(\ref{eq:fusion-etas-temporal-integral}) gives the unstandardized fixed temporal integral used in Eq.~(\ref{eq:fusion-etas-time-feature}); Fusion fixed and Fusion VP then enter the same temporal Fusion representation in Eq.~(\ref{eq:fusion-temporal-feature}) of the main text.

The fixed magnitude branch uses the fixed Gutenberg--Richter parameter $\beta_{3.0}$ while retaining the current input-cutoff scale:
\begin{equation}
F_i^{\mathrm{fixed}}(m)
=
\frac{
1-\exp[-\beta_{3.0}(m-M_{\mathrm{cut}})]
}{
1-\exp[-\beta_{3.0}(M_{\max}-M_{\mathrm{cut}})]
},
\qquad
M_{\mathrm{cut}}\leq m\leq M_{\max}.
\label{eq:fusion-fixed-mag-feature}
\end{equation}
The standardized temporal and magnitude scalars in Eqs.~(\ref{eq:fusion-etas-time-feature}) and~(\ref{eq:fusion-etas-mag-feature}) are the ETAS-derived prior features supplied to the non-linear Fusion module. The target-event likelihood later determines which retained events contribute event log-density terms, without redefining the retained-catalog feature construction above.

\section{Target-event likelihood derivation}
\label{app:target-likelihood}

This note gives the algebra behind the target-event likelihood used in the ``Target-event likelihood and evaluation metrics'' subsection of Methods. The retained catalog is determined by $M_{\mathrm{cut}}$ and provides the conditioning history $H_t$, whereas the forecasting target is the process restricted to events with $m\geq M_d$. For a model $\mathcal{M}$, a retained-catalog marked intensity can be factorized as
\begin{equation}
\lambda^{\mathcal{M}}(t,m\mid H_t)
=
\lambda_t^{\mathcal{M}}(t\mid H_t)
f^{\mathcal{M}}(m\mid t,H_t),
\label{eq:supp-retained-marked-intensity}
\end{equation}
where $f^{\mathcal{M}}$ is normalized on the retained magnitude range. The probability mass assigned to target magnitudes is
\begin{equation}
p_d^{\mathcal{M}}(t\mid H_t)
=
\int_{M_d}^{M_{\max}}
f^{\mathcal{M}}(u\mid t,H_t)\,du .
\label{eq:supp-target-mass}
\end{equation}
Restricting the marked process to the target-event magnitude range gives the temporal target-event intensity
\begin{equation}
\lambda_{t,d}^{\mathcal{M}}(t\mid H_t)
=
\int_{M_d}^{M_{\max}}
\lambda^{\mathcal{M}}(t,u\mid H_t)\,du
=
\lambda_t^{\mathcal{M}}(t\mid H_t)
p_d^{\mathcal{M}}(t\mid H_t),
\label{eq:supp-target-temporal-intensity}
\end{equation}
and the conditional target-event magnitude density
\begin{equation}
f_d^{\mathcal{M}}(m\mid t,H_t)
=
\frac{
f^{\mathcal{M}}(m\mid t,H_t)
}{
p_d^{\mathcal{M}}(t\mid H_t)
},
\qquad
m\in[M_d,M_{\max}].
\label{eq:supp-target-density}
\end{equation}
Equations~(\ref{eq:supp-target-temporal-intensity}) and~(\ref{eq:supp-target-density}) lead to the target-event marked intensity factorization in Eq.~(\ref{eq:rate}) of the main text. ETAS uses this restriction explicitly through the Gutenberg--Richter probability mass. NPP and Fusion are trained and evaluated directly with target-restricted cumulative temporal hazards and target-event magnitude distributions, so $p_d^{\mathcal{M}}$ provides the probabilistic interpretation rather than an additional neural-network output.

Let $i_n$ be the retained-event index of the $n$-th target event. Written over target-event indices, the marked target-event likelihood is
\begin{equation}
\log L_{\mathcal{M}}
=
\sum_n
\left[
\log \lambda_{t,d}^{\mathcal{M}}(t_{i_n}\mid H_{i_n})
+
\log f_d^{\mathcal{M}}(m_{i_n}\mid \Delta_{i_n},H_{i_n})
-
\int_{t_{i_{n-1}}}^{t_{i_n}}
\lambda_{t,d}^{\mathcal{M}}(s\mid H_s)\,ds
\right].
\label{eq:supp-target-index-likelihood}
\end{equation}
The cumulative temporal intensity between two consecutive target events can be decomposed into disjoint retained-event intervals:
\begin{equation}
\int_{t_{i_{n-1}}}^{t_{i_n}}
\lambda_{t,d}^{\mathcal{M}}(s\mid H_s)\,ds
=
\sum_{r=i_{n-1}+1}^{i_n}
\Lambda_{d,r}^{\mathcal{M}}(\Delta_r\mid H_r).
\label{eq:supp-retained-interval-decomposition}
\end{equation}
Substituting Eq.~(\ref{eq:supp-retained-interval-decomposition}) into Eq.~(\ref{eq:supp-target-index-likelihood}) gives the retained-interval objective in	Eq.~(\ref{eq:target-ll}) of the main text. In this form, the event log-intensity and magnitude log-density terms are multiplied by $\mathrm{mask}_i$, whereas the cumulative temporal intensity is accumulated across every retained interval. This is the survival contribution of the target-event process: it accounts for how much target-event intensity the model assigns before the next retained event, regardless of whether that retained event is itself a target event.
The evaluation window terminates at $t_N$, the occurrence time of the final retained test event. The retained-interval objective therefore includes the survival contribution through the final interval ending at $t_N$, with no additional interval or terminal survival contribution after that event.

\section{Target-conditioned ETAS formulation}
\label{app:target-etas}

The fitted ETAS process is defined on the retained catalog determined by $M_{\mathrm{cut}}$, whereas the forecasting target is fixed at $M_d=3.0$. For threshold-specific ETAS fits and Fusion VP, the ETAS productivity term is written relative to the input cutoff:
\begin{equation}
k(m_j)
=
k_0\exp\!\left[a(m_j-M_{\mathrm{cut}})\right],
\label{eq:etas-productivity}
\end{equation}
where $k_0$ controls the productivity scale and $a$ controls the magnitude dependence of triggering productivity. In Fusion fixed, the ETAS parameter set $\theta_{3.0}$ is estimated at $M_{\mathrm{cut}}=3.0$; therefore the productivity component inside $\lambda_{3.0}^{\mathrm{ETAS}}$ uses the same expression with $3.0$ in place of $M_{\mathrm{cut}}$. The temporal triggering kernel follows the normalized Omori--Utsu form
\begin{equation}
g(\tau)
=
\frac{(\omega-1)c^{\omega-1}}{(\tau+c)^{\omega}},
\qquad
\tau>0,
\label{eq:etas-omori}
\end{equation}
where $c$ is the short-time offset parameter and $\omega>1$ controls the aftershock decay rate. The cumulative Omori--Utsu kernel $A_{\mathrm{OU}}(\tau)$ satisfies $A_{\mathrm{OU}}'(\tau)=g(\tau)$ and is
\begin{equation}
A_{\mathrm{OU}}(\tau)
=
\int_0^{\tau}g(u)\,du
=
1-
\left(\frac{c}{\tau+c}\right)^{\omega-1}.
\label{eq:etas-cumulative-kernel}
\end{equation}

For the retained-event interval $(t_{i-1},t_i]$, the fitted ETAS cumulative temporal intensity before target conditioning is
\begin{align}
\Lambda_{0,i}^{\mathrm{ETAS}}
&=
\int_{t_{i-1}}^{t_i}
\lambda_0^{\mathrm{ETAS}}(s\mid H_s)\,ds \notag\\
&=
\mu\Delta_i
+
\sum_{j<i}
k(m_j)
\left[
A_{\mathrm{OU}}(t_i-t_j)
-
A_{\mathrm{OU}}(t_{i-1}-t_j)
\right].
\label{eq:etas-cumulative}
\end{align}
Here $H_s$ follows the same definition as $H_t$, namely the retained history before the integration time $s$. This quantity is accumulated over retained-event intervals. Non-target retained events do not activate event likelihood terms, but they remain in the conditioning history and can affect subsequent triggering calculations.

The ETAS magnitude component follows the truncated Gutenberg--Richter density over $[M_{\mathrm{cut}},M_{\max}]$:
\begin{equation}
f_{\mathrm{ETAS}}(m)
=
\frac{
\beta\exp[-\beta(m-M_{\mathrm{cut}})]
}{
1-\exp[-\beta(M_{\max}-M_{\mathrm{cut}})]
},
\qquad
M_{\mathrm{cut}}\leq m\leq M_{\max},
\label{eq:etas-mag-density}
\end{equation}
with cumulative distribution
\begin{equation}
F_{\mathrm{ETAS}}(m)
=
\frac{
1-\exp[-\beta(m-M_{\mathrm{cut}})]
}{
1-\exp[-\beta(M_{\max}-M_{\mathrm{cut}})]
}.
\label{eq:etas-mag-cdf}
\end{equation}
Let
\begin{equation}
G(m)
=
\int_{M_{\mathrm{cut}}}^{m}
f_{\mathrm{ETAS}}(u)\,du .
\label{eq:Gm}
\end{equation}
The ETAS probability mass assigned to the target-event magnitude range is
\begin{align}
p_d 
&= 
\Pr(M_d\leq M\leq M_{\max}\mid M_{\mathrm{cut}}\leq M\leq M_{\max}) \notag\\
&=
\frac{G(M_{\max})-G(M_d)}
{G(M_{\max})}.
\label{eq:pd}
\end{align}
This probability mass is used for target conditioning: it converts the retained-catalog ETAS process into the target-event process scored at the fixed threshold $M_d$. It is evaluated within the likelihood protocol and is distinct from the cutoff-alignment factor used to construct the Fusion fixed temporal feature in Appendix~\ref{sec:etas-feature-construction}.
The target-conditioned ETAS magnitude density is
\begin{equation}
f_{d}^{\mathrm{ETAS}}(m)
=
\frac{
f_{\mathrm{ETAS}}(m)
}{
G(M_{\max})-G(M_d)
},
\qquad
M_d\leq m\leq M_{\max}.
\label{eq:etas-target-mag-density}
\end{equation}
The corresponding target-event temporal intensity and retained-interval cumulative temporal intensity are
\begin{equation}
\lambda_{t,d}^{\mathrm{ETAS}}(t\mid H_t)
=
p_d\,
\lambda_{0}^{\mathrm{ETAS}}(t\mid H_t),
\label{eq:etas-target-intensity}
\end{equation}
and
\begin{equation}
\Lambda_{d,i}^{\mathrm{ETAS}}
=
p_d\,
\Lambda_{0,i}^{\mathrm{ETAS}}.
\label{eq:etas-target-cumulative}
\end{equation}

\section{Neural architecture and automatic differentiation}

For NPP and Fusion, the recurrent history representation is built from normalized retained-event intervals and magnitudes. The input uses $\log(\Delta_i+\varepsilon_t)$ and $\log m_i$ standardized by training-set statistics, where $\varepsilon_t=10^{-10}$ is added to every elapsed-time value for numerical stability. The recurrent history window length is fixed at $L=20$. The same log-standardization, $[\log(\delta+\varepsilon_t)-\mu_t]/\sigma_t$, is applied to the candidate elapsed time $\delta$ supplied to both the temporal and magnitude branches; no additional clipping or special treatment is used. The cumulative hazard function network (CHFN) outputs the cumulative temporal intensity $\Lambda_{d,i}^{\mathcal{M}}(\delta\mid H_i)$ as a function of $\delta$, and the cumulative magnitude function network (CMFN) outputs the cumulative magnitude distribution $F_{d,i}^{\mathcal{M}}(m\mid \delta,H_i)$ over the target-event magnitude range.

The monotone-network construction follows the NPP implementation of Stockman et al. All trainable kernels on CHFN paths carrying candidate-time dependence, including the direct $\delta$ path and the ETAS temporal-feature path in Fusion, are constrained to be non-negative. Likewise, all kernels on CMFN paths carrying candidate-magnitude dependence, including the direct $m$ path and the Gutenberg--Richter feature path in Fusion, are constrained to be non-negative. The transformations on these paths use non-decreasing activations; conditioning terms that do not depend on the variable being differentiated can shift the network state without changing the derivative sign. CHFN uses a softplus output activation, which makes the cumulative temporal output non-negative, and CMFN uses a sigmoid output activation, which bounds the cumulative magnitude output between zero and one. Because the full paths from $\delta$ to $\Lambda_{d,i}^{\mathcal{M}}$ and from $m$ to $F_{d,i}^{\mathcal{M}}$ are non-decreasing, automatic differentiation yields non-negative temporal intensities and magnitude densities.

The observed event time and magnitude enter only when evaluating the likelihood at $\delta=\Delta_i$ and $m=m_i$. The candidate magnitude input to CMFN is target-centered, $\tilde m=m-M_d$; since $\partial \tilde m/\partial m=1$, automatic differentiation with respect to $m$ gives the same conditional density.

In the Fusion temporal branch, the LSTM history representation, ETAS temporal feature and candidate elapsed time are embedded before concatenation. In the magnitude branch, the main Fusion model excludes the LSTM hidden state and uses the ETAS/Gutenberg--Richter magnitude feature, candidate elapsed time and candidate magnitude input. This branch-specific design is used to preserve the history sensitivity of temporal triggering while avoiding unnecessary sequence-dependent noise in magnitude-density estimation.

\section{Temporal information-gain definitions}
\label{app:information-gain}

Let $i_n$ denote the retained-event index of the $n$-th target event. We set $i_0$ to the retained-event index at the start of the evaluation window, after the conditioning history has been initialized. Thus, for the first target event, the cumulative hazard term includes all retained-event intervals from the evaluation-window start to that target event. The interval between two consecutive target events is
\begin{equation}
\tau_n=t_{i_n}-t_{i_{n-1}}
=
\sum_{r=i_{n-1}+1}^{i_n}\Delta_r .
\label{eq:target-event-interval}
\end{equation}
For two models $A$ and $B$, the event-wise temporal information gain for the target event at time $t_{i_n}$ is
\begin{equation}
\mathrm{TIG}_{n}^{A\mid B}
=
\ell_{\mathrm{temp},n}^{A}
-
\ell_{\mathrm{temp},n}^{B},
\label{eq:event-wise-tig}
\end{equation}
where
\begin{equation}
\ell_{\mathrm{temp},n}^{\mathcal{M}}
=
\log \lambda_{t,d}^{\mathcal{M}}(t_{i_n}\mid H_{i_n})
-
\sum_{r=i_{n-1}+1}^{i_n}
\Lambda_{d,r}^{\mathcal{M}}(\Delta_r\mid H_r).
\label{eq:event-wise-temp-ll}
\end{equation}
Thus, the event-wise contribution contains the target-event log-intensity at $t_{i_n}$ and the cumulative target-event hazard over all retained intervals since the previous target event. The temporal cumulative information gain of Fusion relative to a benchmark model $B$ is
\begin{equation}
\mathrm{TCIG}_{\mathrm{Fusion}\mid B}(t)
=
\sum_{n:t_{i_n}\leq t}
\mathrm{TIG}_{n}^{\mathrm{Fusion}\mid B}.
\label{eq:tcig}
\end{equation}
The positive-event fraction is
\begin{equation}
P_+^{A\mid B}
=
\frac{1}{N_d}
\sum_{n=1}^{N_d}
\mathbb{I}
\left(
\mathrm{TIG}_{n}^{A\mid B}>0
\right)
\times 100\% .
\label{eq:positive-ratio}
\end{equation}

\begin{figure}[!htb]
\centering
\includegraphics[width=\textwidth]{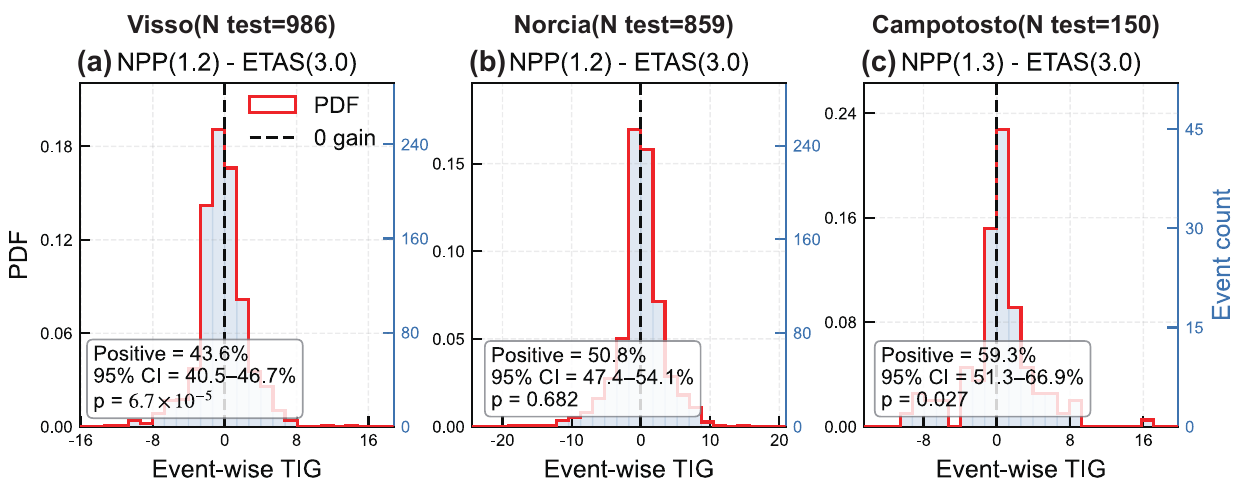}
\caption{\textbf{Event-wise temporal information gain distributions for NPP relative to ETAS on AVN target events.}
Each histogram shows the temporal information gain (TIG) for individual test target events in the Visso, Norcia and Campotosto splits. 
TIG is defined as the event-wise temporal log-likelihood difference between NPP and ETAS, after accounting for the cumulative target-event temporal intensity over all retained intervals since the previous target event. 
\textbf{a--c}, NPP--ETAS comparisons for the Visso, Norcia and Campotosto splits, respectively. 
NPP is evaluated at the low-cutoff settings, \(M_{\mathrm{cut}}=1.2\) for Visso and Norcia and \(M_{\mathrm{cut}}=1.3\) for Campotosto, whereas ETAS is evaluated at the \(M_{\mathrm{cut}}=3.0\) reference setting. 
The dashed vertical line marks zero gain. Positive TIG values indicate target events for which NPP assigns higher temporal likelihood than ETAS. 
Red outlines show the estimated TIG probability density, and the light blue bars show the corresponding event counts. 
The annotation in each panel reports the percentage of events with TIG \(>0\), the corresponding 95\% confidence interval and the nominal two-sided sign-test \(p\) value. 
}
\label{fig:S_tig_PDF_origin}
\end{figure}

As a descriptive significance diagnostic for the positive-event fractions reported in main text Fig.~\ref{fig4:tig_PDF} and Fig.~\ref{fig:S_tig_PDF_origin}, we also use a nominal two-sided sign test against a 50\% event-wise win rate. The test treats the signs of event-wise TIG values as Bernoulli outcomes and evaluates whether the event-wise win rate differs from 50\% under equal event-wise performance. Because target events in earthquake sequences can be temporally correlated, these p values are interpreted as descriptive diagnostics rather than formal independent-event tests. Detailed information is specified in Table~\ref{tab:TIG-value}.

\begin{table}[!htb]
\caption{Event-wise temporal information-gain sign-test statistics. The table reports the number of positive TIG values \(N_+\), the total number of target events \(N_d\), the positive-event fraction, the Wilson 95\% confidence interval (CI), and the two-sided sign-test \(p\) value for the comparisons shown in main text Fig.~\ref{fig4:tig_PDF} and Fig.~\ref{fig:S_tig_PDF_origin}.}
\label{tab:TIG-value}

\centering
\footnotesize
\setlength{\tabcolsep}{4pt}

\begin{tabular}{llcccc}
\hline
Comparison & Split & \(N_+/N_d\) & \shortstack{Positive-event\\fraction} & 95\% CI & \(p\) value \\
\hline
Fusion--ETAS & Visso
& \(541/986\) & \(54.9\%\) & \(51.7\)--\(57.9\%\) & \(0.0025\) \\

Fusion--ETAS & Norcia
& \(502/859\) & \(58.4\%\) & \(55.1\)--\(61.7\%\)
& \(8.5\times10^{-7}\) \\

Fusion--ETAS & Campotosto
& \(86/150\) & \(57.3\%\) & \(49.3\)--\(65.0\%\) & \(0.086\) \\

Fusion--NPP & Visso
& \(638/986\) & \(64.7\%\) & \(61.7\)--\(67.6\%\)
& \(1.8\times10^{-20}\) \\

Fusion--NPP & Norcia
& \(564/859\) & \(65.7\%\) & \(62.4\)--\(68.8\%\)
& \(3.0\times10^{-20}\) \\

Fusion--NPP & Campotosto
& \(71/150\) & \(47.3\%\) & \(39.5\)--\(55.3\%\) & \(0.568\) \\

NPP--ETAS & Visso
& \(430/986\) & \(43.6\%\) & \(40.5\)--\(46.7\%\)
& \(6.7\times10^{-5}\) \\

NPP--ETAS & Norcia
& \(436/859\) & \(50.8\%\) & \(47.4\)--\(54.1\%\) & \(0.682\) \\

NPP--ETAS & Campotosto
& \(89/150\) & \(59.3\%\) & \(51.3\)--\(66.9\%\) & \(0.027\) \\
\hline
\end{tabular}
\end{table}

To further contextualize the Fusion-based event-wise information-gain results in main text Fig.~\ref{fig4:tig_PDF}, we also compare the purely neural NPP baseline directly against the ETAS reference. Figure~\ref{fig:S_tig_PDF_origin} shows the event-wise temporal information gain distributions for NPP--ETAS on the same AVN target-event sequences used in the Fusion comparisons. NPP is evaluated at the low-cutoff settings, \(M_{\mathrm{cut}}=1.2\) for the Visso and Norcia splits and \(M_{\mathrm{cut}}=1.3\) for the Campotosto split, whereas ETAS is evaluated at the \(M_{\mathrm{cut}}=3.0\) reference setting.

The NPP--ETAS distributions show that the purely neural baseline does not consistently outperform the ETAS reference at the event level. In the Visso split, only 43.6\% of target events have positive TIG, meaning that NPP assigns higher temporal likelihood than ETAS to fewer than half of the target events. In the Norcia split, the positive fraction is 50.8\%, close to random balance. The Campotosto split shows a higher positive fraction of 59.3\%, but this result is based on a much smaller target-event set (\(N_{\mathrm{test}}=150\)) and should therefore be interpreted cautiously. Compared with the Fusion--ETAS results in main text Fig.~\ref{fig4:tig_PDF}a--c, these diagnostics are consistent with the ETAS-derived temporal feature contributing to Fusion's event-wise gains beyond the use of a neural point-process architecture alone.

\section{Spatial distributions of benchmark earthquake catalogs in California}
\label{app:catalog-domains}

\begin{figure}[!htb]
\centering
\includegraphics[width=\textwidth]{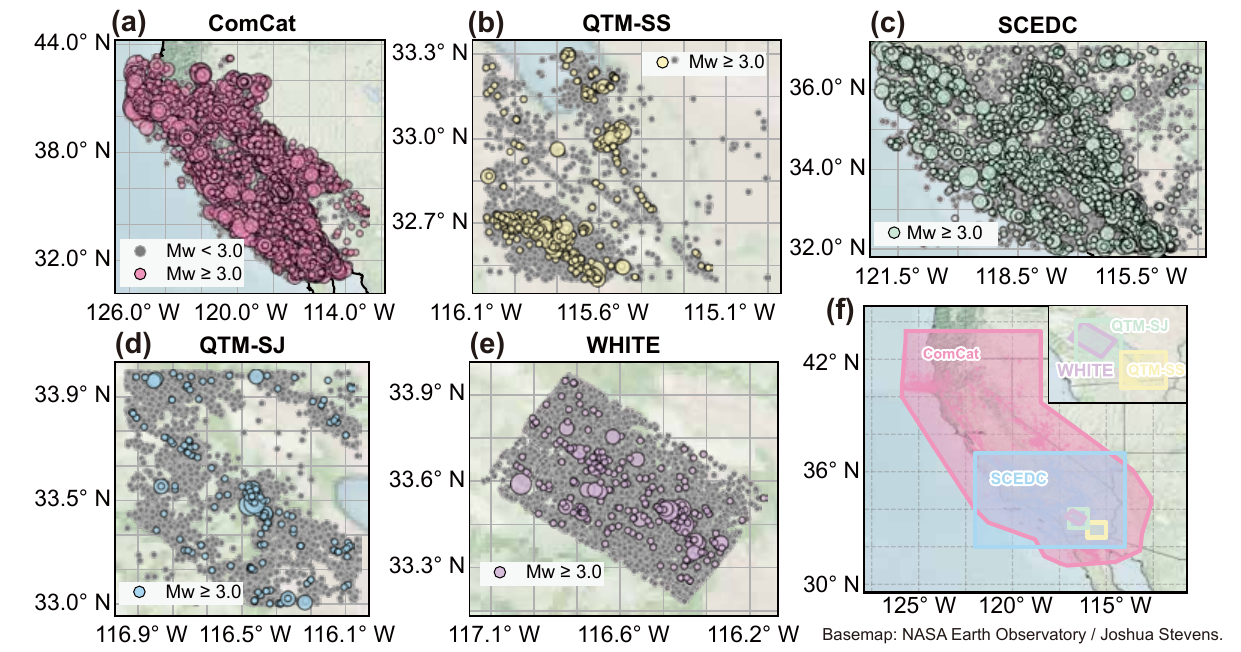}
\caption{\textbf{Spatial distributions of the EarthquakeNPP benchmark catalogs.}
\textbf{a--e}, Epicentral distributions of ComCat, QTM-SaltonSea (QTM-SS), SCEDC, QTM-SanJacinto (QTM-SJ), and WHITE, respectively. Gray markers denote events with catalog-reported magnitude \(m<3.0\), whereas colored markers denote target events with \(m\geq3.0\). Marker size is scaled by catalog-reported magnitude.
\textbf{f}, Combined spatial overview of the five catalogs, with the inset enlarging the localized WHITE, QTM-SS, and QTM-SJ regions. All likelihood evaluations use \(M_d=3.0\), while the catalog-specific \(M_{\mathrm{cut}}\) determines the retained input history. Catalog periods, event counts, and minimum input cutoffs are summarized in Table~\ref{tab:benchmark-spatial-domain}.}
\label{fig:S_benchmark_spatial}
\end{figure}

Figure~\ref{fig:S_benchmark_spatial} and Table~\ref{tab:benchmark-spatial-domain} summarize the spatial domains and catalog metadata of the five EarthquakeNPP benchmark catalogs. They differ in spatial coverage, duration, event count and benchmark-specified minimum input cutoff. ComCat spans California broadly, whereas SCEDC, QTM-SS, QTM-SJ and WHITE cover more localized southern California regions.

\begin{table}[!htb]
\caption{\textbf{Catalog metadata and spatial domains of the EarthquakeNPP benchmark datasets.} 
The table summarizes the benchmark-specified minimum input cutoff, catalog period, event count and longitude--latitude range for each
benchmark catalog.}
\label{tab:benchmark-spatial-domain}

\centering
\footnotesize
\setlength{\tabcolsep}{4pt}
\renewcommand{\arraystretch}{1.15}

\begin{tabular}{lccc p{7.2cm}}
\hline
Dataset
& Period
& Min. \(M_{\mathrm{cut}}\)
& \(N\)
& Spatial domain \\
\hline

ComCat
& 1971--2020
& 2.5
& 92,102
& \(125.899^\circ\,\mathrm{W}\)--\(112.705^\circ\,\mathrm{W}\),
  \(31.067^\circ\,\mathrm{N}\)--\(43.363^\circ\,\mathrm{N}\) \\

QTM-SS
& 2008--2017
& 1.0
& 45,570
& \(116.000^\circ\,\mathrm{W}\)--\(115.004^\circ\,\mathrm{W}\),
  \(32.500^\circ\,\mathrm{N}\)--\(33.302^\circ\,\mathrm{N}\) \\

QTM-SJ
& 2008--2017
& 1.0
& 21,291
& \(117.000^\circ\,\mathrm{W}\)--\(115.999^\circ\,\mathrm{W}\),
  \(33.000^\circ\,\mathrm{N}\)--\(34.001^\circ\,\mathrm{N}\) \\

SCEDC
& 1981--2020
& 2.0
& 130,357
& \(122.002^\circ\,\mathrm{W}\)--\(114.000^\circ\,\mathrm{W}\),
  \(32.000^\circ\,\mathrm{N}\)--\(37.002^\circ\,\mathrm{N}\) \\

WHITE
& 2008--2021
& 0.6
& 58,636
& \(117.124^\circ\,\mathrm{W}\)--\(116.153^\circ\,\mathrm{W}\),
  \(33.183^\circ\,\mathrm{N}\)--\(33.965^\circ\,\mathrm{N}\) \\
\hline
\end{tabular}
\end{table}

All benchmark likelihoods use the common target threshold \(M_d=3.0\). NPP and Fusion use each dataset's minimum \(M_{\mathrm{cut}}\) in the reported model-specific comparison, while ETAS uses the magnitude-complete \(M_{\mathrm{cut}}=3.0\) reference. For the five California catalogs, Fusion uses one ETAS parameter set estimated from the ComCat Auxiliary and Training periods at \(M_{\mathrm{cut}}=3.0\), applied to the \(m\geq3.0\) ETAS feature history of the catalog being evaluated. This shared feature source tests performance consistency across heterogeneous catalogs under the specified protocol rather than complete regional transferability.

\section{Fusion-m configuration and AVN log-likelihood results}

In addition to the main Fusion framework described in main text Fig.~\ref{fig1:framework}c, we also evaluated an alternative configuration, denoted as Fusion-m. Fusion-m follows the same catalog preprocessing, target-event definition and likelihood-based evaluation protocol as the main Fusion model. The main difference lies in the construction of the magnitude branch. In the main Fusion framework, the magnitude branch does not directly use the LSTM hidden representation; instead, it combines the candidate elapsed time, the candidate magnitude input and the ETAS/Gutenberg--Richter-derived magnitude feature. By contrast, Fusion-m additionally introduces the LSTM-based history representation into the magnitude branch. Thus, the magnitude feature in Fusion-m is constructed from the LSTM output, the ETAS-derived magnitude feature, the candidate elapsed time and the candidate magnitude input. This configuration is used as an ablation to examine the effect of adding recurrent-history information directly to the magnitude branch.

\begin{figure}[!htb]
\centering
\includegraphics[width=\textwidth]{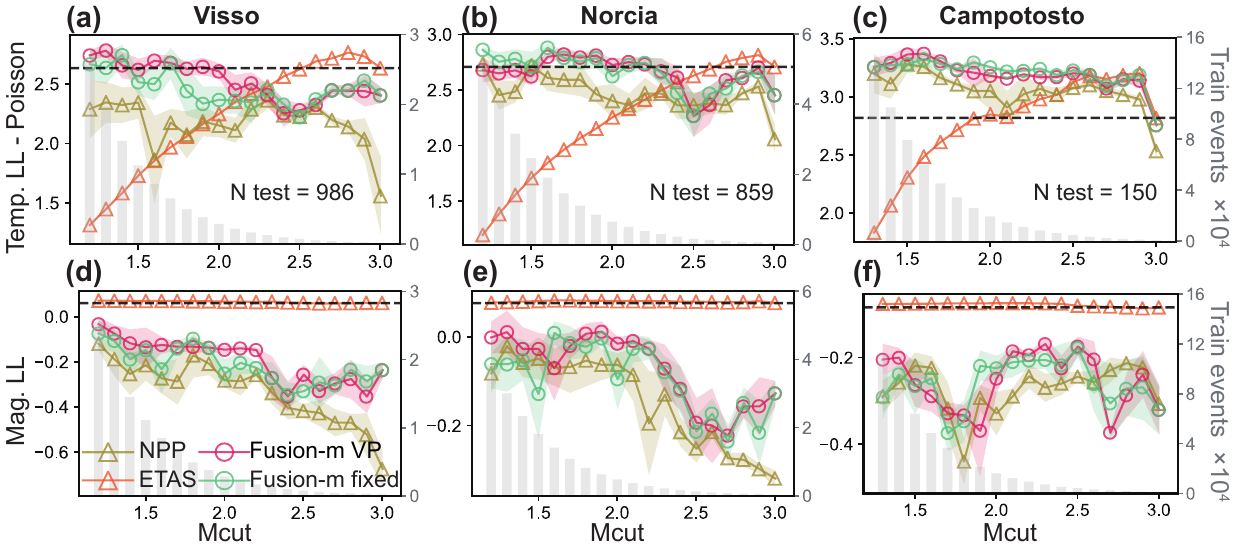}
\caption{\textbf{Fusion-m log-likelihood comparison across input catalog cutoffs on the AVN earthquake catalog.}
\textbf{a--c}, Temporal log-likelihood relative to the Poisson benchmark for the Visso, Norcia and Campotosto splits, respectively. 
\textbf{d--f}, Magnitude log-likelihood for the same splits, reported on the original likelihood scale. 
Fusion-m is an ablation in which the magnitude branch additionally receives the LSTM hidden representation, unlike the main Fusion model. 
The compared models are ETAS, NPP, Fusion-m VP and Fusion-m fixed across different retained-catalog cutoffs $M_{\mathrm{cut}}$, with all likelihoods evaluated on target events above $M_d=3.0$. 
Shaded bands indicate variability across repeated random initializations for neural-network-based models; ETAS is fitted deterministically and is shown without an initialization-dependent band. 
Gray bars show the number of training events under each $M_{\mathrm{cut}}$, scaled by the right-hand axis in units of $10^4$. 
The annotated $N_{\mathrm{test}}$ values give the number of target events in each test split. 
Black dashed lines show the ETAS reference obtained from the fixed $M_{\mathrm{cut}}=3.0$ parameter setting.}
\label{fig:S_LL}
\end{figure}

Figure~\ref{fig:S_LL} reports the log-likelihood results of Fusion-m on the AVN catalog across different input cutoff thresholds $M_{\mathrm{cut}}$. The presentation follows the same format as main text Fig.~\ref{fig2:LL}: panels \textbf{a--c} show temporal log-likelihood relative to the homogeneous Poisson benchmark, whereas panels \textbf{d--f} show magnitude log-likelihood on the original scale. Fusion-m has higher temporal likelihood than NPP in several cutoff settings. Its magnitude likelihood is generally lower and more variable than that of the main Fusion model in main text Fig.~\ref{fig2:LL}, showing that direct recurrent-history input does not consistently improve the magnitude branch in these experiments.

The comparison between Fig.~\ref{fig:S_LL} and main text Fig.~\ref{fig2:LL} is consistent with the branch-specific design used in the main Fusion framework. The recurrent representation remains available to the temporal branch, where recent retained history can inform timing, while the magnitude branch is informed by the Gutenberg--Richter-derived feature without direct LSTM input. The ablation therefore supports the more limited conclusion in the main text: recurrent history does not provide a consistent additional benefit for target-event magnitude-density estimation.

\section{Additional temporal-intensity comparison}

\begin{figure}[!htb]
\centering
\includegraphics[width=\textwidth]{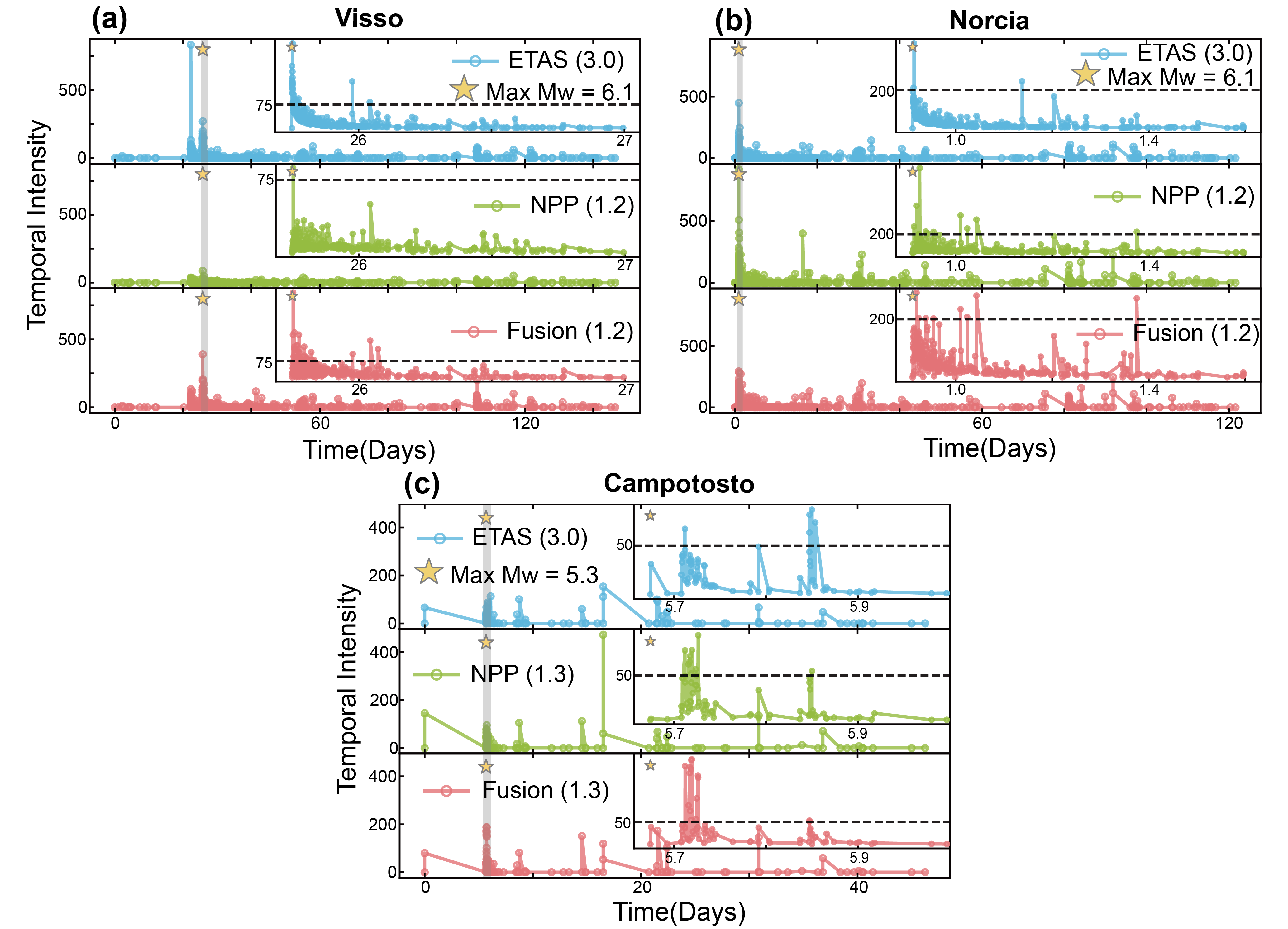}
\caption{\textbf{Temporal-intensity comparison among ETAS, NPP and Fusion on the AVN earthquake sequence.}
\textbf{a--c}, Model-estimated target-event temporal intensities for the Visso, Norcia and Campotosto splits, respectively, with ETAS, NPP and Fusion shown from top to bottom. Insets enlarge the gray-shaded intervals, and stars mark the largest-magnitude test event. ETAS uses $M_{\mathrm{cut}}=3.0$; NPP and Fusion use $M_{\mathrm{cut}}=1.2$ for Visso and Norcia and $M_{\mathrm{cut}}=1.3$ for Campotosto. All intensities refer to target events with $m\geq M_d=3.0$.}
\label{fig:S_temp_rate}
\end{figure}

To complement the log-likelihood, TCIG and event-wise TIG analyses in the main text, we compare the estimated temporal intensities of ETAS, NPP and the main Fusion model. This analysis corresponds to the Fusion architecture shown in main text Fig.~\ref{fig1:framework}c, whose temporal branch combines the LSTM history representation with the ETAS-derived temporal feature. 
Figure~\ref{fig:S_temp_rate} is a descriptive visualization of selected model configurations rather than an independent model-comparison experiment.

Figure~\ref{fig:S_temp_rate} visualizes the model-estimated temporal intensity on the AVN test sequences. The displayed configurations are selected from representative settings with strong temporal log-likelihood performance: ETAS uses the fixed $M_{\mathrm{cut}}=3.0$ setting, whereas NPP and Fusion use the lower-cutoff settings that provide richer historical information, namely $M_{\mathrm{cut}}=1.2$ for the Visso and Norcia splits and $M_{\mathrm{cut}}=1.3$ for the Campotosto split. This setting allows the temporal-rate structures of the three models to be compared under representative configurations used in the main AVN analysis.

ETAS displays the smooth parametric response implied by its Omori--Utsu kernel, while NPP produces several sharper local peaks. Fusion maintains elevated intensity across multiple aftershock-active intervals and is less dominated by isolated peaks in the displayed configurations. These qualitative patterns are consistent with the TCIG and event-wise TIG results in main text Figs.~\ref{fig3:TCIG} and ~\ref{fig4:tig_PDF}, where gains accumulate through substantial portions of the Visso and Norcia sequences. The intensity plots do not identify a unique mechanism, but show how the likelihood differences are distributed in time for the selected configurations.

\section{Magnitude log-likelihood of benchmark earthquake catalogs in California}
\label{app:magnitude-results}

\begin{figure}[!htb]
\centering
\includegraphics[width=\textwidth]{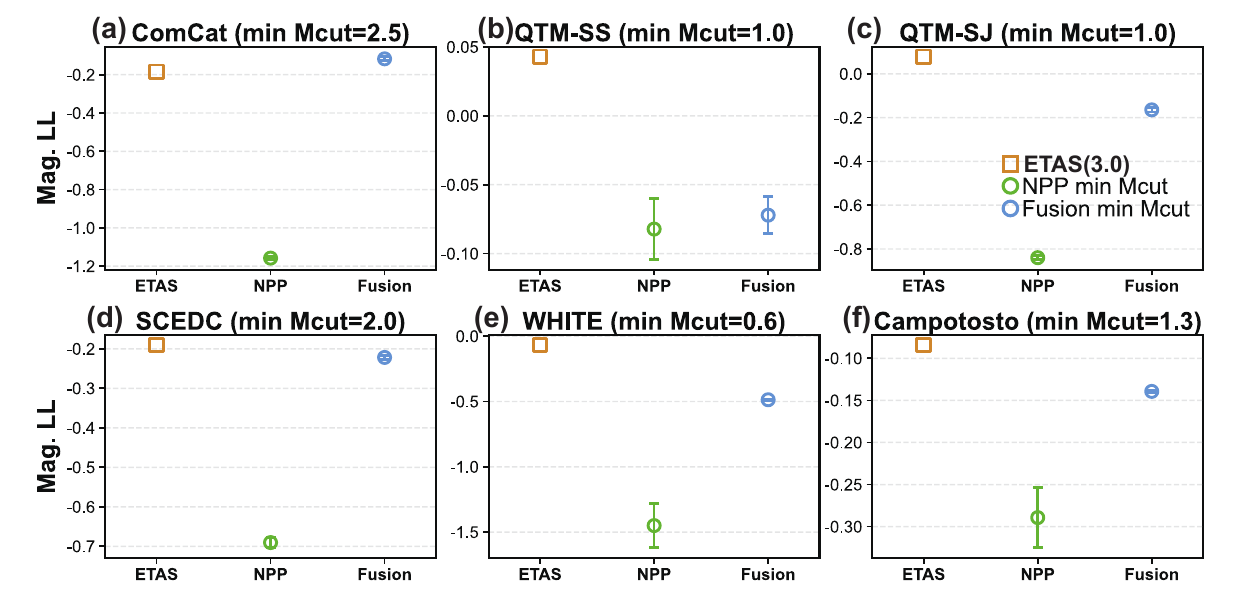}
\caption{\textbf{Target-event magnitude log-likelihood comparison across the EarthquakeNPP benchmark catalogs and the Campotosto AVN split.}
Panels \textbf{a--f} correspond to ComCat, QTM-SaltonSea (QTM-SS), QTM-SanJacinto (QTM-SJ), SCEDC, WHITE and Campotosto, respectively. NPP and Fusion use the dataset-specific minimum \(M_{\mathrm{cut}}\), whereas ETAS uses the \(M_{\mathrm{cut}}=3.0\) reference. Markers and error bars show the mean \(\pm\) one standard deviation over repeated random initializations for NPP and Fusion; ETAS is deterministic. The California Fusion experiments share ETAS parameters estimated from the ComCat training history at \(M_{\mathrm{cut}}=3.0\), while Campotosto uses the AVN fixed-parameter configuration. All likelihoods use the common target threshold \(M_d=3.0\); higher values indicate better target-event magnitude-density estimation.}
\label{fig:S_benchmark_box_mag}
\end{figure}

To evaluate the scaling-law-informed magnitude branch across additional catalogs, Fig.~\ref{fig:S_benchmark_box_mag} presents target-event magnitude log-likelihood for the five EarthquakeNPP benchmark catalogs and the Campotosto AVN split. The magnitude evaluation uses the same catalog partitions, target threshold and model-specific input configurations as the temporal comparison.

For NPP and Fusion, the results use the dataset-specific minimum \(M_{\mathrm{cut}}\) shown above each panel; ETAS uses the \(M_{\mathrm{cut}}=3.0\) reference. Neural-model markers show the mean over repeated random initializations, with error bars indicating one standard deviation, whereas ETAS is deterministic. All magnitude likelihoods are evaluated on the same target events satisfying \(m\geq M_d=3.0\), where \(m\) denotes catalog-reported magnitude.

\begin{table}[!htb]
\caption{\textbf{Temporal and magnitude log-likelihood results for the EarthquakeNPP benchmark catalogs and the Campotosto AVN split.}
Results for the five EarthquakeNPP benchmark catalogs correspond to Fig.~\ref{fig5:benchmark_box} of the main text and Fig.~\ref{fig:S_benchmark_box_mag}, whereas the Campotosto results correspond to Figs.~\ref{fig2:LL}c and~\ref{fig2:LL}f of the main text. Results are reported under the dataset-specific minimum \(M_{\mathrm{cut}}\) and \(M_{\mathrm{cut}}=3.0\); ETAS is shown only for \(M_{\mathrm{cut}}=3.0\). Temp.\ LL and Mag.\ LL denote target-event temporal and magnitude log-likelihood, respectively. Bold values indicate the best result within each dataset and likelihood component.}
\label{tab:benchmark-likelihood-results}

\centering
\footnotesize
\setlength{\tabcolsep}{2.5pt}
\renewcommand{\arraystretch}{1.15}

\begin{tabular}{llcccc}
\hline
Dataset
& Model
& \multicolumn{2}{c}{Minimum \(M_{\mathrm{cut}}\)}
& \multicolumn{2}{c}{\(M_{\mathrm{cut}}=3.0\)} \\
\cline{3-4}\cline{5-6}
&
& Temp.\ LL
& Mag.\ LL
& Temp.\ LL
& Mag.\ LL \\
\hline

{\shortstack[c]{ComCat\\(2.5)}}
& ETAS
& --
& --
& \(-2.4091\)
& \(-0.1844\) \\

& NPP
& \(-2.3815 \pm 0.0039\)
& \(-1.1579 \pm 0.0073\)
& \(-2.3977 \pm 0.0133\)
& \(-1.1533 \pm 0.0032\) \\

& Fusion
& \(\mathbf{-2.3538 \pm 0.0118}\)
& \(-0.1171 \pm 0.0110\)
& \(-2.4385 \pm 0.0642\)
& \(\mathbf{-0.0597 \pm 0.0464}\) \\
\hline

{\shortstack[c]{QTM-SS\\(1.0)}}
& ETAS
& --
& --
& \(-6.7590\)
& \(\mathbf{0.0429}\) \\

& NPP
& \(-4.5061 \pm 0.1071\)
& \(-0.0822 \pm 0.0221\)
& \(-5.1582 \pm 0.0642\)
& \(-0.3916 \pm 0.0490\) \\

& Fusion
& \(\mathbf{-4.3013 \pm 0.0364}\)
& \(-0.0721 \pm 0.0133\)
& \(-5.1867 \pm 0.1451\)
& \(-0.1873 \pm 0.0225\) \\
\hline

{\shortstack[c]{QTM-SJ\\(1.0)}}
& ETAS
& --
& --
& \(-9.3168\)
& \(\mathbf{0.0782}\) \\

& NPP
& \(-6.0370 \pm 0.1597\)
& \(-0.8397 \pm 0.0084\)
& \(-6.5164 \pm 0.2002\)
& \(-2.2434 \pm 0.4829\) \\

& Fusion
& \(\mathbf{-5.8521 \pm 0.1100}\)
& \(-0.1644 \pm 0.0099\)
& \(-6.4848 \pm 0.2625\)
& \(-0.3709 \pm 0.0649\) \\
\hline

\shortstack{SCEDC\\(2.0)}
& ETAS
& --
& --
& \(-1.3314\)
& \(\mathbf{-0.1907}\) \\

& NPP
& \(-1.3902 \pm 0.0119\)
& \(-0.6906 \pm 0.0136\)
& \(-1.3110 \pm 0.0233\)
& \(-0.6661 \pm 0.0125\) \\

& Fusion
& \(\mathbf{-1.1627 \pm 0.0086}\)
& \(-0.2218 \pm 0.0045\)
& \(-1.3838 \pm 0.0928\)
& \(-0.1955 \pm 0.0009\) \\
\hline

{\shortstack[c]{WHITE\\(0.6)}}
& ETAS
& --
& --
& \(-7.8313\)
& \(\mathbf{-0.0686}\) \\

& NPP
& \(-6.2545 \pm 0.0498\)
& \(-1.4502 \pm 0.1714\)
& \(-6.3803 \pm 0.0513\)
& \(-1.8911 \pm 0.1372\) \\

& Fusion
& \(-6.1629 \pm 0.0492\)
& \(-0.4868 \pm 0.0045\)
& \(\mathbf{-6.1441 \pm 0.0584}\)
& \(-0.4272 \pm 0.0848\) \\
\hline

{\shortstack[c]{Campotosto\\(1.3)}}
& ETAS
& --
& --
& \(-0.3643\)
& \(\mathbf{-0.0844}\) \\

& NPP
& \(-0.2779 \pm 0.1667\)
& \(-0.2892 \pm 0.0361\)
& \(-0.6503 \pm 0.0524\)
& \(-0.3065 \pm 0.0368\) \\

& Fusion
& \(\mathbf{-0.2088 \pm 0.0376}\)
& \(-0.1393 \pm 0.0016\)
& \(-0.4724 \pm 0.0103\)
& \(-0.3254 \pm 0.0135\) \\
\hline
\end{tabular}
\end{table}

Fusion has higher magnitude log-likelihood than NPP in all five benchmark catalogs and the Campotosto split. ComCat is the only panel in which Fusion also exceeds the Gutenberg--Richter-based ETAS reference. This catalog-specific result does not extend to the other benchmarks and is therefore not interpreted as a general advantage over the parametric magnitude model.

ETAS achieves the highest magnitude log-likelihood for QTM-SS, QTM-SJ, SCEDC, WHITE and Campotosto, whereas Fusion is highest for ComCat. This pattern is consistent with main text Figs.~\ref{fig2:LL}d--f: the scaling-law-informed magnitude branch narrows the gap between NPP and the Gutenberg--Richter-based ETAS reference but does not systematically exceed that reference.

The cross-catalog conclusion therefore matches the main text: the Gutenberg--Richter-derived feature improves the neural magnitude branch relative to NPP under the evaluated configurations, while the ETAS/Gutenberg--Richter distribution remains the stronger overall magnitude reference.

\noappendix

\codedataavailability{
The AVN dataset used in Figs.~\ref{fig1:framework}a, b, ~\ref{fig2:LL}--~\ref{fig4:tig_PDF} and ~\ref{fig5:benchmark_box}f was originally developed by ~\citet{ref72} and is publicly available at \url{https://doi.org/10.5281/zenodo.4662870}~\citep{Tan2021AVNData}. The processed version used in this study is available in the Neural Point Process repository at \url{https://github.com/ss15859/Neural-Point-Process}~\citep{ref44}. The five EarthquakeNPP benchmark datasets used in Figs.~\ref{fig5:benchmark_box}a--e are publicly available at \url{https://github.com/ss15859/EarthquakeNPP}~\citep{ref69}. We gratefully acknowledge the original catalog developers, the USGS, SCEDC and SCSN for producing and maintaining these earthquake data resources. The source code is available at \url{https://github.com/XiongTLu/FusionEarthquake}.} 

\authorcontribution{TX: methodology, software, data curation, visualization,
writing -- original draft, and writing -- review and editing. ZZ: software
and visualization. YL: software and visualization. WL: supervision.
YA: funding acquisition, supervision, and writing -- review and editing.
YZ: methodology, project administration, funding acquisition, supervision,
and writing -- review and editing.} 

\competinginterests{The authors declare that they have no conflict of interest.} 


\begin{acknowledgements}
The authors gratefully acknowledge Tan et al. for developing and publicly releasing the Amatrice--Visso--Norcia earthquake dataset. 
We also thank the original catalog developers, the U.S. Geological Survey, the Southern California Earthquake Data Center, and the Southern California Seismic Network for producing and maintaining the earthquake data resources used in this study.
The authors used ChatGPT (OpenAI) to check and polish the manuscript language and to assist with code review and writing portions of the code. 
All AI-generated code was reviewed and double-checked by the authors.
\end{acknowledgements}

\financialsupport{
This research was supported by the National Natural Science Foundation of China (Young Scientists Fund, grant no. GZ2023GC020A; Key Program, grant no. ZZ2023QW001A) and the Yunnan Revitalization Talents Support Plan of China (grant no. CB26138T019A).
}


\bibliographystyle{copernicus}
\bibliography{references}

@article{ref08,
  author = {Chen, Min and Qian, Zhen and Boers, Niklas and Creutzig, Felix and Camps-Valls, Gustau and Hubacek, Klaus and Claramunt, Christophe and Wilson, John P. and Nativi, Stefano and Jakeman, Anthony J. and M{\"u}ller, R. Dietmar and Batty, Michael and Zhou, Chenghu and Chen, Fahu and Wang, Qiao and Zhang, Fan and Barton, C. Michael and Strobl, Josef and Meadows, Michael and Ratti, Carlo and Hess, Philipp and Xu, Qingsong and Zhang, Zhixin and Gu, Qiushi and Zhu, A-Xing and Lin, Hui and Yuan, Linwang and L{\"u}, Guonian},
  title = {{Collaboration between artificial intelligence and Earth science communities for mutual benefit}},
  journal = {Nature Geoscience},
  year = {2024},
  volume = {17},
  number = {10},
  pages = {949--952},
  doi = {10.1038/s41561-024-01550-x}
}

@article{ref09,
  author = {Chen, Min and Qian, Zhen and Boers, Niklas and Jakeman, Anthony J. and Kettner, Albert J. and Brandt, Martin and Kwan, Mei-Po and Batty, Michael and Li, Wenwen and Zhu, Rui and Luo, Wei and Ames, Daniel P. and Barton, C. Michael and Cuddy, Susan M. and Koirala, Sujan and Zhang, Fan and Ratti, Carlo and Liu, Jian and Zhong, Teng and Liu, Junzhi and Wen, Yongning and Yue, Songshan and Zhu, Zhiyi and Zhang, Zhixin and Sun, Zhuo and Lin, Jian and Ma, Zaiyang and He, Yuanqing and Xu, Kai and Zhang, Chunxiao and Lin, Hui and L{\"u}, Guonian},
  title = {{Iterative integration of deep learning in hybrid Earth surface system modelling}},
  journal = {Nature Reviews Earth and Environment},
  year = {2023},
  volume = {4},
  number = {8},
  pages = {568--581},
  doi = {10.1038/s43017-023-00452-7}
}

@article{ref10,
  author = {Dascher-Cousineau, Kelian and Shchur, Oleksandr and Brodsky, Emily E. and G{\"u}nnemann, Stephan},
  title = {{Using Deep Learning for Flexible and Scalable Earthquake Forecasting}},
  journal = {Geophysical Research Letters},
  year = {2023},
  volume = {50},
  number = {17},
  eid = {e2023GL103909},
  doi = {10.1029/2023GL103909}
}

@article{ref11,
  author = {Dramsch, Jesper S{\"o}ren and Kuglitsch, Monique M. and Fern{\'a}ndez-Torres, Miguel-{\'A}ngel and Toreti, Andrea and Albayrak, Rustem Arif and Nava, Lorenzo and Ghaffarian, Saman and Cheng, Ximeng and Ma, Jackie and Samek, Wojciech and Venguswamy, Rudy and Koul, Anirudh and Muthuregunathan, Raghavan and Hrast Essenfelder, Arthur},
  title = {{Explainability can foster trust in artificial intelligence in geoscience}},
  journal = {Nature Geoscience},
  year = {2025},
  volume = {18},
  number = {2},
  pages = {112--114},
  doi = {10.1038/s41561-025-01639-x}
}

@article{ref12,
  author = {Gelbrecht, Maximilian and White, Alistair and Bathiany, Sebastian and Boers, Niklas},
  title = {{Differentiable programming for Earth system modeling}},
  journal = {Geoscientific Model Development},
  year = {2023},
  volume = {16},
  number = {11},
  pages = {3123--3135},
  doi = {10.5194/gmd-16-3123-2023}
}

@article{ref16,
  author = {Gulia, L. and Wiemer, S.},
  title = {{Real-time discrimination of earthquake foreshocks and aftershocks}},
  journal = {Nature},
  year = {2019},
  volume = {574},
  number = {7777},
  pages = {193--199},
  doi = {10.1038/s41586-019-1606-4}
}

@article{ref17,
  author = {Gutenberg, B. and Richter, C. F.},
  title = {{Frequency of earthquakes in California}},
  journal = {Bulletin of the Seismological Society of America},
  year = {1944},
  volume = {34},
  pages = {185--188},
  doi = {10.1785/BSSA0340040185},
}

@article{ref19,
  author = {Hardebeck, Jeanne L. and Llenos, Andrea L. and Michael, Andrew J. and Page, Morgan T. and Schneider, Max and van der Elst, Nicholas J.},
  title = {{Aftershock Forecasting}},
  journal = {Annual Review of Earth and Planetary Sciences},
  year = {2024},
  volume = {52},
  number = {1},
  pages = {61--84},
  doi = {10.1146/annurev-earth-040522-102129}
}

@article{ref22,
  author = {Hu, Yumeng and Zhang, Qi and Zhu, Hengshu and Wang, Baoshan and Xiong, Hui and Wang, Haitao},
  title = {{Scalable intermediate-term earthquake forecasting with multimodal fusion neural networks}},
  journal = {Scientific Reports},
  year = {2025},
  volume = {15},
  number = {1},
  pages = {9748},
  doi = {10.1038/s41598-025-93877-7}
}

@article{ref25,
  author = {Irrgang, Christopher and Boers, Niklas and Sonnewald, Maike and Barnes, Elizabeth A. and Kadow, Christopher and Staneva, Joanna and Saynisch-Wagner, Jan},
  title = {{Towards neural Earth system modelling by integrating artificial intelligence in Earth system science}},
  journal = {Nature Machine Intelligence},
  year = {2021},
  volume = {3},
  number = {8},
  pages = {667--674},
  doi = {10.1038/s42256-021-00374-3}
}

@article{ref26,
  author = {Iwata, D. and Nanjo, K. Z.},
  title = {{Adaptive estimation of the Gutenberg--Richter b value using a state space model and particle filtering}},
  journal = {Scientific Reports},
  year = {2024},
  volume = {14},
  number = {1},
  pages = {4630},
  doi = {10.1038/s41598-024-54576-x}
}

@article{ref27,
  author = {Jordan, Thomas H. and Chen, Yun-Tai and Gasparini, Paolo and Madariaga, Raul and Main, Ian and Marzocchi, Warner and Papadopoulos, Gerassimos and Sobolev, Gennady and Yamaoka, Koshun and Zschau, Jochen},
  title = {{Operational earthquake forecasting: State of knowledge and guidelines for utilization}},
  journal = {Annals of Geophysics},
  year = {2011},
  volume = {54},
  number = {4},
  pages = {315--391},
  doi = {10.4401/ag-5350}
}

@article{ref28,
  author = {Karniadakis, George Em and Kevrekidis, Ioannis G. and Lu, Lu and Perdikaris, Paris and Wang, Sifan and Yang, Liu},
  title = {{Physics-informed machine learning}},
  journal = {Nature Reviews Physics},
  year = {2021},
  volume = {3},
  number = {6},
  pages = {422--440},
  doi = {10.1038/s42254-021-00314-5}
}

@article{ref31,
  author = {Marzocchi, W. and Taroni, M. and Falcone, G.},
  title = {{Earthquake forecasting during the complex Amatrice-Norcia seismic sequence}},
  journal = {Science Advances},
  year = {2017},
  volume = {3},
  number = {9},
  eid = {e1701239},
  doi = {10.1126/sciadv.1701239}
}

@article{ref32,
  author = {Mizrahi, Leila and Dallo, Irina and van der Elst, Nicholas J. and Christophersen, Annemarie and Spassiani, Ilaria and Werner, Maximilian J. and Iturrieta, Pablo and Bayona, Jos{\'e} A. and Iervolino, Iunio and Schneider, Max and Page, Morgan T. and Zhuang, Jiancang and Herrmann, Marcus and Michael, Andrew J. and Falcone, Giuseppe and Marzocchi, Warner and Rhoades, David and Gerstenberger, Matt and Gulia, Laura and Schorlemmer, Danijel and Becker, Julia and Han, Marta and Kuratle, Lorena and Marti, Mich{\`e}le and Wiemer, Stefan},
  title = {{Developing, Testing, and Communicating Earthquake Forecasts: Current Practices and Future Directions}},
  journal = {Reviews of Geophysics},
  year = {2024},
  volume = {62},
  number = {3},
  eid = {e2023RG000823},
  doi = {10.1029/2023RG000823}
}

@article{ref33,
  author = {Mizrahi, L. and Nandan, S. and Wiemer, S.},
  title = {{Embracing Data Incompleteness for Better Earthquake Forecasting}},
  journal = {Journal of Geophysical Research: Solid Earth},
  year = {2021},
  volume = {126},
  number = {12},
  eid = {e2021JB022379},
  doi = {10.1029/2021JB022379}
}

@article{ref35,
  author = {Mousavi, S. M. and Cattania, C. and Beroza, G. C.},
  title = {{The pursuit of reliable earthquake forecasting}},
  journal = {Physics Today},
  year = {2025},
  volume = {78},
  number = {8},
  pages = {42--50},
  doi = {10.1063/pt.tdwq.nppx}
}

@article{ref36,
  author = {Ogata, Y.},
  title = {{Statistical models for earthquake occurrences and residual analysis for point processes}},
  journal = {Journal of the American Statistical Association},
  year = {1988},
  volume = {83},
  number = {401},
  pages = {9--27},
  doi = {10.1080/01621459.1988.10478560}
}

@article{ref37,
  author = {Ogata, Y.},
  title = {{Space-time point-process models for earthquake occurrences}},
  journal = {Annals of the Institute of Statistical Mathematics},
  year = {1998},
  volume = {50},
  number = {2},
  pages = {379--402},
  doi = {10.1023/A:1003403601725}
}

@article{ref38,
  author = {Ogata, Y.},
  title = {{Significant improvements of the space-time ETAS model for forecasting of accurate baseline seismicity}},
  journal = {Earth, Planets and Space},
  year = {2011},
  volume = {63},
  number = {3},
  pages = {217--229},
  doi = {10.5047/eps.2010.09.001}
}

@article{ref39,
  author = {Ross, Zachary E. and Trugman, Daniel T. and Hauksson, Egill and Shearer, Peter M.},
  title = {{Searching for hidden earthquakes in Southern California}},
  journal = {Science},
  year = {2019},
  volume = {364},
  number = {6442},
  pages = {767--771},
  doi = {10.1126/science.aaw6888}
}

@article{ref41,
  author = {Schultz, R. and Wiemer, S.},
  title = {{Forecasting the rate of induced seismicity as a neural temporal point process}},
  journal = {Journal of Geophysical Research: Machine Learning and Computation},
  year = {2026},
  volume = {3},
  number = {1},
  eid = {e2025JH001052},
  doi = {10.1029/2025JH001052}
}

@article{ref42,
  author = {Seif, Stefanie and Mignan, Arnaud and Zechar, Jeremy Douglas and Werner, Maximilian Jonas and Wiemer, Stefan},
  title = {{Estimating ETAS: The effects of truncation, missing data, and model assumptions}},
  journal = {Journal of Geophysical Research: Solid Earth},
  year = {2017},
  volume = {122},
  number = {1},
  pages = {449--469},
  doi = {10.1002/2016JB012809}
}

@article{ref44,
  author = {Stockman, S. and Lawson, D. J. and Werner, M. J.},
  title = {{Forecasting the 2016--2017 Central Apennines Earthquake Sequence With a Neural Point Process}},
  journal = {Earth's Future},
  year = {2023},
  volume = {11},
  number = {9},
  eid = {e2023EF003777},
  doi = {10.1029/2023EF003777}
}

@article{ref47,
  author = {Utsu, T.},
  title = {{A statistical study on the occurrence of aftershocks}},
  journal = {Geophysical Magazine},
  year = {1961},
  volume = {30},
  pages = {521--605}
}

@article{ref48,
  author = {Utsu, T. and Ogata, Y. and Matsu'ura, Ritsuko S.},
  title = {{The centenary of the Omori formula for a decay law of aftershock activity}},
  journal = {Journal of Physics of the Earth},
  year = {1995},
  volume = {43},
  number = {1},
  pages = {1--33},
  doi = {10.4294/jpe1952.43.1}
}

@article{ref49,
  author = {Wang, Hanchen and Fu, Tianfan and Du, Yuanqi and Gao, Wenhao and Huang, Kexin and Liu, Ziming and Chandak, Payal and Liu, Shengchao and Van Katwyk, Peter and Deac, Andreea and Anandkumar, Anima and Bergen, Karianne and Gomes, Carla P. and Ho, Shirley and Kohli, Pushmeet and Lasenby, Joan and Leskovec, Jure and Liu, Tie-Yan and Manrai, Arjun and Marks, Debora and Ramsundar, Bharath and Song, Le and Sun, Jimeng and Tang, Jian and Veli{\v{c}}kovi{\'c}, Petar and Welling, Max and Zhang, Linfeng and Coley, Connor W. and Bengio, Yoshua and Zitnik, Marinka},
  title = {{Scientific discovery in the age of artificial intelligence}},
  journal = {Nature},
  year = {2023},
  volume = {620},
  number = {7972},
  pages = {47--60},
  doi = {10.1038/s41586-023-06221-2}
}

@article{ref50,
  author = {Wang, Xinyi and Li, Jiawei and Feng, Ao and Sornette, Didier},
  title = {{Estimating Magnitude Completeness in Earthquake Catalogs: A Comparative Study of Catalog-Based Methods}},
  journal = {Journal of Geophysical Research: Solid Earth},
  year = {2025},
  volume = {130},
  number = {9},
  eid = {e2025JB031441},
  doi = {10.1029/2025JB031441}
}

@article{ref53,
  author = {Zhan, Chengxiang and Gao, Shichen and Zhang, Ying and Li, Jiawei and Meng, Qingyan},
  title = {{ETAS-Inspired Spatio-Temporal Convolutional (STC) Model for Next-Day Earthquake Forecasting}},
  journal = {IEEE Transactions on Geoscience and Remote Sensing},
  year = {2024},
  volume = {62},
  pages = {1--14},
  doi = {10.1109/TGRS.2024.3424881}
}

@article{ref54,
  author = {Zhang, Haoyuan and Ke, Shuya and Liu, Wenqi and Zhang, Yongwen},
  title = {{A combining earthquake forecasting model between deep learning and epidemic-type aftershock sequence (ETAS) model}},
  journal = {Geophysical Journal International},
  year = {2024},
  volume = {239},
  number = {3},
  eid = {1545--1556},
  doi = {10.1093/gji/ggae349}
}

@article{ref55,
  author = {Zhang, Ying and Wen, Congcong and Zhan, Chengxiang and Sornette, Didier},
  title = {{Integrating artificial intelligence and geophysical insights for earthquake forecasting: A cross-disciplinary review}},
  journal = {Earth-Science Reviews},
  year = {2025},
  volume = {270},
  eid = {105232},
  doi = {10.1016/j.earscirev.2025.105232}
}

@article{ref56,
  author = {Zhang, Ying and Zhan, Chengxiang and Huang, Qinghua and Sornette, Didier},
  title = {{Forecasting future earthquakes with deep neural networks: Application to California}},
  journal = {Geophysical Journal International},
  year = {2025},
  volume = {240},
  number = {1},
  pages = {81--95},
  doi = {10.1093/gji/ggae373}
}

@article{ref57,
  author = {Zlydenko, Oleg and Elidan, Gal and Hassidim, Avinatan and Kukliansky, Doron and Matias, Yossi and Meade, Brendan and Molchanov, Alexandra and Nevo, Sella and Bar-Sinai, Yohai},
  title = {{A neural encoder for earthquake rate forecasting}},
  journal = {Scientific Reports},
  year = {2023},
  volume = {13},
  number = {1},
  eid = {12350},
  doi = {10.1038/s41598-023-38033-9}
}

@article{ref58,
  author = {Zhang, Yongwen and Fan, Jingfang and Marzocchi, Warner and Shapira, Avi and Hofstetter, Rami and Havlin, Shlomo and Ashkenazy, Yosef},
  title = {{Scaling laws in earthquake memory for interevent times and distances}},
  journal = {Physical Review Research},
  year = {2020},
  volume = {2},
  number = {1},
  eid = {013264},
  doi = {10.1103/PhysRevResearch.2.013264}
}

@article{ref59,
  author = {Zhang, Yongwen and Zhou, Dong and Fan, Jingfang and Marzocchi, Warner and Ashkenazy, Yosef and Havlin, Shlomo},
  title = {{Improved earthquake aftershocks forecasting model based on long-term memory}},
  journal = {New Journal of Physics},
  year = {2021},
  volume = {23},
  number = {4},
  eid = {042001},
  doi = {10.1088/1367-2630/abeb46}
}

@article{ref60,
  author = {Zhang, Yongwen and Ashkenazy, Yosef and Havlin, Shlomo},
  title = {{Asymmetry in Earthquake Interevent Time Intervals}},
  journal = {Journal of Geophysical Research: Solid Earth},
  year = {2021},
  volume = {126},
  number = {9},
  eid = {e2021JB022454},
  doi = {10.1029/2021JB022454}
}

@article{ref63,
  author = {Zhang, Ying and Zhan, Chengxiang and Huang, Qinghua and Sornette, Didier},
  title = {{Seismically Informed Reference Models Enhance AI-Based Earthquake Prediction Systems}},
  journal = {Journal of Geophysical Research: Solid Earth},
  year = {2024},
  volume = {129},
  number = {3},
  eid = {e2023JB028037},
  doi = {10.1029/2023JB028037}
}

@article{ref65,
  author = {Li, Jiawei and Sornette, Didier and Wu, Zhongliang and Zhuang, Jiancang and Jiang, Changsheng},
  title = {{Revisiting seismicity criticality: A new framework for bias correction of statistical seismology model calibrations}},
  journal = {Journal of Geophysical Research: Solid Earth},
  year = {2025},
  volume = {130},
  number = {6},
  eid = {e2024JB029337},
  doi = {10.1029/2024JB029337}
}

@article{ref69,
  title = {{Earthquake{NPP}: A Benchmark for Earthquake Forecasting with Neural Point Processes}},
  author = {Samuel Stockman and Daniel John Lawson and Maximilian J. Werner},
  journal = {Transactions on Machine Learning Research},
  issn = {2835-8856},
  year = {2026},
  url = {https://openreview.net/forum?id=dIcNAg6ZuZ}
}

@article{ref70,

author = {Nandan, Shyam and Ouillon, Guy and Sornette, Didier and Wiemer, Stefan},

title = {{Forecasting the Rates of Future Aftershocks of All Generations Is Essential to Develop Better Earthquake Forecast Models}},

journal = {Journal of Geophysical Research: Solid Earth},

year = {2019},

volume = {124},

number = {8},

pages = {8404--8425},

doi = {10.1029/2018JB016668}

}

@article{ref71,

author = {Ebrahimian, Hossein and Jalayer, Fatemeh},

title = {{Robust seismicity forecasting based on Bayesian parameter estimation for epidemiological spatio-temporal aftershock clustering models}},

journal = {Scientific Reports},

year = {2017},

volume = {7},

eid = {9803},

doi = {10.1038/s41598-017-09962-z}

}

@article{ref72,
  title={Machine-learning-based high-resolution earthquake catalog reveals how complex fault structures were activated during the 2016--2017 central Italy sequence},
  author={Tan, Yen Joe and Waldhauser, Felix and Ellsworth, William L and Zhang, Miao and Zhu, Weiqiang and Michele, Maddalena and Chiaraluce, Lauro and Beroza, Gregory C and Segou, Margarita},
  journal={The Seismic Record},
  volume={1},
  number={1},
  pages={11--19},
  year={2021},
  publisher={Seismological Society of America},
  doi = {10.1785/0320210001},
}

@misc{ref73,
  author       = {{U.S. Geological Survey}},
  title        = {{Advanced National Seismic System (ANSS) Comprehensive Catalog of Earthquake Events and Products}},
  howpublished = {U.S. Geological Survey, Earthquake Hazards Program [data set]},
  doi          = {10.5066/F7MS3QZH},
  url          = {https://doi.org/10.5066/F7MS3QZH},
  year = {2017},
}

@misc{ref74,
  author       = {{Southern California Earthquake Data Center}},
  title        = {{Southern California Earthquake Center}},
  year         = {2013},
  howpublished = {California Institute of Technology [data set]},
  doi          = {10.7909/C3WD3xH1},
  url          = {https://doi.org/10.7909/C3WD3xH1},
}

@misc{ref75,
  author       = {{California Institute of Technology}},
  title        = {{Southern California Seismic Network}},
  year         = {1926},
  howpublished = {International Federation of Digital Seismograph Networks [seismic network]},
  doi          = {10.7914/SN/CI},
  url          = {https://doi.org/10.7914/SN/CI},
}

@article{ref77,
  author  = {White, Michael C. A. and Ben-Zion, Yehuda and Vernon, Frank L.},
  title   = {A Detailed Earthquake Catalog for the San Jacinto Fault-Zone Region in Southern California},
  journal = {Journal of Geophysical Research: Solid Earth},
  year    = {2019},
  volume  = {124},
  number  = {7},
  pages   = {6908--6930},
  doi     = {10.1029/2019JB017641},
  url     = {https://doi.org/10.1029/2019JB017641}
}

@misc{Tan2021AVNData,
  author       = {Tan, Yen Joe and Waldhauser, Felix and Ellsworth, William},
  title        = {{Machine-Learning-Based High-Resolution Earthquake Catalog For the 2016--2017 Central Italy Sequence}},
  year         = {2021},
  howpublished = {Zenodo [data set]},
  doi          = {10.5281/zenodo.4662870},
  url          = {https://doi.org/10.5281/zenodo.4662870}
}

\end{document}